%% file: sc24.tex
\documentclass[conference]{IEEEtran}
\input{packages}
\IEEEoverridecommandlockouts
\UseRawInputEncoding

\newcommand{\SysName}{\textsc{TorchGT}\xspace}
\newcommand{\RawGT}{\textsc{GP-Raw}\xspace}
\newcommand{\FlashGT}{\textsc{GP-Flash}\xspace}
\newcommand{\SparseGT}{\textsc{GP-Sparse}\xspace}

\usepackage{hyperref}
\hypersetup{
  colorlinks   = true,    
  urlcolor     = olive,    
  linkcolor    = blue,    
  citecolor    = blue      
}

\begin{document}



\title{\SysName: A Holistic System for Large-scale Graph Transformer Training}



\author{
\IEEEauthorblockN{Meng Zhang\textsuperscript{\IEEEauthorrefmark{1}1,3} 
    \quad Jie Sun\textsuperscript{\IEEEauthorrefmark{1}4} 
    \quad Qinghao Hu\textsuperscript{1,3} 
    \quad Peng Sun\textsuperscript{3,5} 
    \quad Zeke Wang\textsuperscript{4} 
    \quad Yonggang Wen\textsuperscript{2} 
    \\
    Tianwei Zhang\textsuperscript{\thanks{\IEEEauthorrefmark{1}Equal Contribution.  \textsuperscript{\ding{73}}Corresponding author.}\ding{73}2}
    }
\vspace{5pt}
\IEEEauthorblockA{
    \textsuperscript{1}S-Lab, Nanyang Technological University 
    \qquad \textsuperscript{2}NTU
    \qquad \textsuperscript{3}Shanghai AI Laboratory
    \\
    \textsuperscript{4}Zhejiang University
    \qquad \textsuperscript{5}SenseTime Research
    }
}








\maketitle 
\pagestyle{plain} 

\input{0_Abstract}
\input{1_Introduction}

\input{2_Background_and_Motivation}

\input{3_System_Design}
\input{4_Evaluation}

\input{5_Related_Work}

\input{6_Conclusion}


\section*{Acknowledgments}
We sincerely thank our anonymous SC reviewers for their valuable comments. This research is supported under the National Key
R\&D Program of China (2022ZD0160201) and the RIE2020 Industry Alignment Fund - Industry Collaboration Projects (IAF-ICP) Funding Initiative, as well as cash and in-kind contribution from the industry partner(s).

\bibliographystyle{IEEEtran}
\bibliography{sc24}

\end{document}

%% file: packages.tex
\usepackage{xspace}
\usepackage{mathtools} 
\usepackage{amssymb}
\usepackage{bm}
\usepackage{enumitem}
\usepackage{cite}
\usepackage{textcomp}
\usepackage{xcolor}
\usepackage{changepage}

\usepackage{graphicx}
\usepackage{subcaption}
\usepackage{tikz}

\usepackage{tabularx}
\usepackage{multirow}
\usepackage{booktabs}
\usepackage{makecell}
\usepackage[figuresleft]{rotating}
\usepackage[flushleft]{threeparttable}

\usepackage{algorithmicx}
\usepackage{algorithm}
\usepackage[noend]{algpseudocode}

\algnewcommand\algorithmicinput{\textbf{Input:}}
\algnewcommand\Input{\item[\algorithmicinput]}
\algnewcommand\algorithmicoutput{\textbf{Output:}}
\algnewcommand\Output{\item[\algorithmicoutput]}

\usepackage[utf8]{inputenc} 
\usepackage{tcolorbox} 
\usepackage{soul} 
\usepackage{siunitx} 
\usepackage{pifont} 
\usepackage{bbding} 
\usepackage[hidelinks]{hyperref} 
\usepackage[switch]{lineno} 

\newcommand{\one}{\raisebox{-0.6mm}{\large{\ding{182}}}}
\newcommand{\two}{\raisebox{-0.6mm}{\large{\ding{183}}}}

\usepackage[skip=3pt]{caption}

%% file: 0_Abstract.tex
\begin{abstract}
Graph Transformer is a new architecture that surpasses GNNs in graph learning. While there emerge inspiring algorithm advancements, their practical adoption is still limited, particularly on real-world graphs involving up to millions of nodes. We observe existing graph transformers fail on large-scale graphs mainly due to heavy computation, limited scalability and inferior model quality. 

Motivated by these observations, we propose \SysName, the \textit{first} efficient, scalable, and accurate graph transformer training system. \SysName optimizes training at three different levels. At \textit{algorithm} level, by harnessing the graph sparsity, \SysName introduces a Dual-interleaved Attention which is computation-efficient and accuracy-maintained. At \textit{runtime} level, \SysName scales training across workers with a communication-light Cluster-aware Graph Parallelism. At \textit{kernel} level, an Elastic Computation Reformation further optimizes the computation by reducing memory access latency in a dynamic way. Extensive experiments demonstrate that \SysName boosts training by up to 62.7$\times$ and supports graph sequence lengths of up to 1M.

\end{abstract}

%% file: 1_Introduction.tex
\section{Introduction}
\label{sec_intro}

Graph-structured data has long been prevalent and indispensable in many real-life applications such as social network construction and molecule analysis. Thus, there emerges a specific family of graph learning methods, namely graph neural networks (GNNs) \cite{GCN,GraphSAGE,GAT}. GNNs have gained giant breakthroughs and exhibit impressive performance in many tasks such as node classification \cite{PowerfulGNN,PUFFIN,CoGNN,Sylvie} and link prediction \cite{GAT}, mainly due to their message passing mechanism \cite{GNN-survey}, which models the inherent properties of graph structures. However, this module in classic GNNs also leads to commonly acknowledged over-smoothing \cite{GNNOverSmooth}, over-squashing \cite{GNNOverSquash,GNNOverSquash22} and limited expressivity \cite{HighOrderGNN} issues. 

To address these deficiencies, a latest approach called \emph{graph transformer} shows more promising power in capturing the inter-dependencies among nodes. Graph transformer is built upon the classical Transformer \cite{Attention} which allows nodes to attend to all other nodes, and integrates multiple graph structure encodings to include important graph properties. Due to the great modeling capability, graph transformer has garnered surging interest in recent years and a large number of models have been proposed \cite{Graphormer,GT,NAGphormer,NodeFormer}. Existing graph transformers mainly operate by treating graph nodes as input tokens and constructing an input sequence consisting of all the graph nodes. Besides, graph structure encoders are designed as a graph adaptation of the original Transformer architecture. By integrating structural information, graph transformers exhibit competitive performance and outperform traditional message-passing GNNs (e.g., GCN \cite{GCN} and GAT \cite{GAT}) on both node classification \cite{NodeFormer,NAGphormer,SAN,SAT,DIFFormer,SGFormer} and graph classification \cite{Graphormer,GT,GraphTrans} tasks, as shown by Table \ref{table_gtasneed}. We can obviously see graph transformers obtain the highest scores than GNNs on all tasks.

\begin{table}[t]
    \centering
    \caption{Graph transformers outperform classical GNNs on graph-level and node-level (Flickr) tasks.}
    \resizebox{\linewidth}{!}{
        \begin{tabular}{@{}ccccc@{}}
\toprule
\multicolumn{2}{c}{\textbf{Model}}                                                                          & \textbf{\begin{tabular}[c]{@{}c@{}}ZINC\\      Test MAE$\downarrow$\end{tabular}} & \textbf{\begin{tabular}[c]{@{}c@{}}PCQM4M-LSC\\      Validate MAE$\downarrow$\end{tabular}} & \textbf{\begin{tabular}[c]{@{}c@{}}Flickr\\      Test Acc.(\%)$\uparrow$\end{tabular}} \\ \midrule
\multirow{2}{*}{\textbf{\begin{tabular}[c]{@{}c@{}}Traditional   \\      GNNs\end{tabular}}}   & GAT        & 0.384                                                                             & -                                                                                           & 54.29                                                                                  \\
                                                                                               & GCN        & 0.367                                                                             & 0.169                                                                                       & 61.49                                                                                  \\ \midrule
\multirow{2}{*}{\textbf{\begin{tabular}[c]{@{}c@{}}Graph   \\      Transformers\end{tabular}}} & GT         & 0.226                                                                             & 0.141                                                                                       & \textbf{68.59}                                                                                  \\
                                                                                               & Graphormer & \textbf{0.122}                                                                             & \textbf{0.123}                                                                                       & 66.16                                                                                  \\ \bottomrule
\end{tabular}
}
\label{table_gtasneed}
\end{table}

Real-world graphs can easily involve millions of nodes \cite{OGB,amazon}, making the sequence length enormously large. For example, in the current graph transformers' operation way, processing the citation graph dataset ogbn-papers100M from Open Graph Benchmark \cite{OGB} (including more than 100 million nodes) requires high dimensional inputs with prohibited sequences. Moreover, as illustrated by the profiled results in \S \ref{subsec_longs}, 
training graph transformers with long sequence is crucial for model quality and the development of versatile graph transformer application scenarios. However, we find most existing graph transformer research works \cite{Graphormer,GT,GraphGPS,Exphormer} are only limited to small graphs due to a lack of compatible systems tailored for the graph transformer model training with long sequences. More specifically, there are three deficiencies in current works:


\textbf{First, graph transformers with standard attention have poor scalability to long sequences}, due to the computation and memory complexity of $O(N^2)$, quadratic on the number of nodes ($N$) in a graph~\cite{Graphormer,GT,smallgraph-case1,smallgraph-case2,NI-CTR,Equiformer}.
Taking fully-connected attention as graph foundation encoders captures the implicit all-pair influence beyond neighboring nodes, but also limits existing graph transformers
only on small-graph applications \cite{Graphormer,GT,smallgraph-case1,smallgraph-case2,NI-CTR,Equiformer}. Figure~\ref{fig_motivation_flashattn} demonstrates that even with a state-of-the-art attention library, i.e., FlashAttention \cite{FlashAttention}, the computation of the dense attention mechanism is still a bottleneck during the graph transformer training.

\textbf{Second, current algorithms either compromise model quality or are only applicable to a single graph learning task}. To reduce computation pressure, some graph transformers shorten the input sequences by harnessing neighbor sampling \cite{NAGphormer,Gophormer} similar adopted in classic GNNs~\cite{FastGCN,GraphSAGE,GraphSAINT}. Others like \cite{GraphGPS} attempt to overcome the quadratic complexity by replacing full attention with approximate attention methods. However, using sampling methods or simply adopting sparse patterns like \cite{Performer,BIGBIRD} loses critical connectivity information and thus sacrifices model precision. On the other hand, some works \cite{NodeFormer,SGFormer,DIFFormer} use self-defined adapted attention modules to reduce memory consumption, but are limited to a specific task, e.g., node classification. They are neither general to versatile graph learning tasks nor portable to be scaled in large-scale training.

\textbf{Third, no existing works exploit systematic optimizations to realize efficient and scalable training}. Several graph transformers \cite{Exphormer,GraphGPS} apply graph structure to relieve the computation burden. However, this sparse pattern is highly irregular in memory access due to the skewed property of graphs, which is challenging for optimizing the system throughput.
Moreover, with large datasets, the memory consumption of model activations grows rapidly, necessitating a scalable system design and memory optimization. But existing works \cite{Graphormer,GT,GraphGPS,Exphormer,NAGphormer} only focus on the implementation of graph transformers in a single GPU, thus limited to very small graphs. Although there has been a breakthrough for large language models (LLMs) by partitioning along the input sequence dimension and training long sequences across devices \cite{sp-colossalai,sp-ds,sp-megatron,BlockwiseRingAttn}, those sequence parallelism ways cannot be directly transplanted on graph transformers due to the extra graph encodings and neglection of structure properties.

To bridge these gaps, we design \SysName, the \textit{\textbf{first}} distributed training system that scales \textit{graph transformer model} to large graphs with billions of edges. Our system abides four design goals: \textit{\textbf{scalable}}, \textit{\textbf{efficient}}, \textit{\textbf{convergence-maintained}}, and \textit{\textbf{task-agnostic}}. Existing graph transformer works neither facilitate efficiency by well-designed parallelism from the system perspective nor propose scalable algorithms for universal graph learning tasks, thus making it challenging to meet those goals. This hinders the practical development of advanced graph transformer models on real-world graphs. 
The core design of \SysName derives from the following three key insights. First, \textit{the learning of graph transformers highly benefits from graph structures.} Specifically, the structure of many real-world graphs is highly sparse \cite{Mizan,Enterprise,Powerlaw}, which reflects the inherent vertex-vertex interactions. This sparsity could be a guide for how graph transformers attend to nodes to reduce computation costs while maintaining correct connections. In addition, considering the structural property in the system design also contributes to optimal hardware throughput. Second, \textit{the order of input graph tokens is alterable.} Unlike inputs in famous LLMs like GPT \cite{GPT3} whose token order is crucial for model understanding and generation process\cite{LLaMA,GPT3}, graph transformers focus more on connections between nodes. Thus, we can modify the input arrangement to exploit graph properties (e.g., local clusters) for more specialized optimizations. Third, \textit{the block-sparse format is a good match for irregular graph clusters.} Block-sparse formats store data contiguously in memory, reducing storage overheads and memory access. But directly exploiting it on dense attention matrices will drop connectivity and result in substantial accuracy loss \cite{BlocksparseRNN,BlockBert}. However, by integrating it into our specialized clustered pattern, we find the computation can be further accelerated while maintaining model accuracy. 

As such, our key idea is to design an accuracy-maintained and compute-efficient system from both algorithm and system perspectives to support large-scale graph transformer training. Specifically, \SysName consists of three key innovations.
\textbf{Dual-interleaved Attention} is a local-global interleaved attention that integrates graph structural topology into the attention module and selectively combines the global information into the attention with the graph structure search, which efficiently speeds up the attention computation while maintaining the models' qualities. \textbf{Cluster-aware Graph Parallelism} splits the input graph tokens according to the cluster nature of graphs, thus boosting the attention computation throughput and facilitating system scalability. It also allows us to take advantage of the cluster feature in more fine-grained kernel optimizations. Inspired by the block-sparse format, \textbf{Elastic Computation Reformation} dynamically transfers the clustered attention pattern into a specialized cluster-sparse format to reduce the irregular memory access latency. It includes an \textit{Auto Tuner} to automatically control the transfer to maintain the model convergence. Through extensive experiments, we show \SysName successfully achieves scalable and efficient graph transformer training on large graphs. It also boosts training by up to 62.7$\times$ across various graph learning tasks while maintaining accuracy.

In summary, we make the following contributions:

\begin{itemize}[leftmargin=*, itemsep=0pt, topsep=0pt, label=\ding{72}]
\item \SysName is the \textit{first} graph transformer system that facilitates efficient, scalable, and accurate training on large-scale graphs as well as universal graph learning tasks.

\item \SysName is the \textit{first} to identify the major challenges that hinder existing graph transformers from scaling to large graphs and explore the graph-specific optimization opportunities which are neglected previously.

\item We propose three key techniques to meet all design goals from algorithm and system co-design perspectives.

\item Experiments show \SysName achieves up to 62.7$\times$ speedup and near-linear scalability, supporting graph sequence lengths of up to millions.


\end{itemize}

%% file: 2_Background_and_Motivation.tex
\section{Background and Motivation}
\label{sec_motivation}

\subsection{Graph Transformer}
\emph{Graph transformer} architecture has attracted surging attention in graph representation learning in recent years \cite{AutoGT}. Current representative graph transformers integrate graph structural encodings into the input and attention map in the Transformer architecture. The input sequence is built by tokens generated with graph attributes. Specifically, some works\cite{GT,Graphormer,NAGphormer,EGT,GraphBert,SAN,UniMP} calculate node positional encodings beforehand and add them to the inputs before the attention module. Other works \cite{Graphormer,GT,Gophormer,NodeFormer,PLAN} add graph structural information into the self-attention matrix as bias. Several works \cite{GraphGPS,SAT,Exphormer} combine message-passing GNNs and the attention mechanism together. Here we only focus on the former two types of graph transformers since they are currently most representative. 

A basic Transformer consists of multi-head attention (MHA) and feed-forward network (FFN) which contains two linear layers. Given an input sequence $\boldsymbol{H}=[h_1,\cdots,h_S]^{\top} \in \mathbb{R}^{S\times d}$ where $S$ is the sequence length and $d$ is the hidden dimension, MHA first projects its input $\boldsymbol{H}$ to three subspaces: $\boldsymbol{Q}$, $\boldsymbol{K}$ and $\boldsymbol{V}$ with projection weight matrices $\boldsymbol{W_Q}\in \mathbb{R}^{d\times d_{\boldsymbol{K}}},\boldsymbol{W_K}\in \mathbb{R}^{d\times d_{\boldsymbol{K}}},\boldsymbol{W_V}\in \mathbb{R}^{d\times d_{\boldsymbol{V}}}$. The MHA output is calculated as:
\begin{equation}
\boldsymbol{H}^{\prime} = \text{softmax}\left(\frac{\boldsymbol{Q}\boldsymbol{K}^{\top}}{\sqrt{d_{\boldsymbol{K}}}}\right)\boldsymbol{V} \label{attncore}
\end{equation}
where $d_{\boldsymbol{K}}$ is the second dimension of $\boldsymbol{K}$. MHA captures the pair-wise similarity of input tokens in the sequence. 

For a graph $G=(V, E)$ with nodes $V=\left\{v_1, \cdots, v_{N}\right\}$ and edges $E=\left\{e_1, \cdots, e_{|E|}\right\}$, here we list the formulation of Graphormer \cite{Graphormer} as an example:
\begin{align}
h_i^{(0)}=x_i+z_{\operatorname{deg}^{-}\left(v_i\right)}^{-}+z_{\operatorname{deg}^{+}\left(v_i\right)}^{+} \label{equ_graphormer1} \\
    \boldsymbol{A}_{i j}=\frac{\left(h_i \boldsymbol{W_Q}\right)\left(h_j \boldsymbol{W_K}\right)^{\top}}{\sqrt{d_{\boldsymbol{K}}}}+bias_{\phi\left(v_i, v_j\right)} \label{equ_graphormer2}
\end{align}
where $h_i^{(0)}$ is the beginning attribute of node $i$, $x_i$ is the node feature, and $z^-,z^+\in \mathbb{R}^{d}$ are learnable embeddings specified by the in-degree $\text{deg}^-(v_i)$ and out-degree $\text{deg}^+(v_i)$. The encodings in Equation \ref{equ_graphormer1} allow the attention to capture the node importance. $\boldsymbol{A}_{i j}$ is the ($i$, $j$)-element of Query-Key product matrix, namely the attention coefficient. $bias_{\phi}$ is a learnable scalar shared across all layers, and $\phi\left(v_i, v_j\right)$ is the distance of the shortest path (SPD) between node $v_i$ and $v_j$, which is the shortest hops that $v_i$ needs to pass to reach $v_j$.

\begin{figure}[t]
    \centering
    \includegraphics[width=\linewidth]{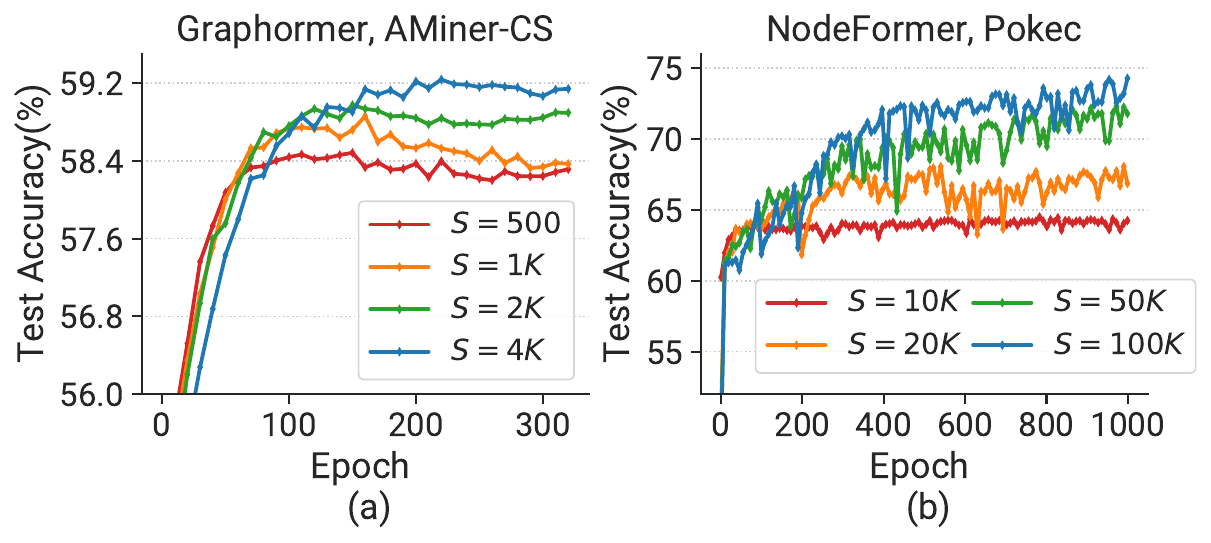}
    \caption{The test accuracy of graph transformers when trained with different sequence lengths $S$.}
    \label{fig_motivation2.2}
\end{figure}

\subsection{Long Sequence for Graph Transformers}
\label{subsec_longs}
For better illustration, we categorize current graph learning tasks into two types to discuss the need of training in long sequences: (1) graph-level task, and (2) node-level task.

\noindent\textbf{Long Sequence for Graph-level Tasks.} 
For such tasks, the input sequences represent a set of graphs while the output is a set of labels representing the types of corresponding graphs. When processed by graph transformers, all nodes in each input graph need to be encoded as input tokens and are concatenated into an input sequence. As such, the length of each sequence equals the number of nodes in each graph. In this task, if the graph size, i.e., the number of nodes, grows very large, the input sequence can be too long to be trained by current methods. For instance, the MalNet~\cite{Malnet} dataset contains graphs with up to 552K nodes. 

\noindent\textbf{Long Sequence for Node-level Tasks.} These tasks classify each node in an input graph with a specific label. In the node classification task, the input sequences can either encode all nodes in the graph or a mini-batch of nodes. For the former case, the input sequence can be enormously long for large-scale graphs, which is not supported by most models. For the latter, with a larger batch size, both the training throughput and the trained model quality can be improved. Both cases validate the necessity and advantages of long sequence training. 

However, existing graph transformers have some inherent constraints in performing the above tasks. While graph-level scenario has been explored in \cite{Graphormer,GT}, existing endeavors do not generalize to large-scale graphs endemic to node-level prediction. Our \SysName strives to include both tasks by joint algorithm-system design. The scale of graphs applicable to current models is still limited, thus leaving long sequence training still an urgent necessity. Besides, training large-scale graphs in short sequence suffers from lower training throughput, downgraded model quality and limited graph transformer applications.
Figure \ref{fig_motivation2.2} illustrates the impact of sequence length on the test accuracy of two representative models Graphormer \cite{Graphormer} and NodeFormer \cite{NodeFormer} on two datasets. Both models show superior performance on longer sequences. On the AMiner-CS dataset, Graphormer with a 4K sequence length improves the test accuracy by up to 0.9\% compared to the short sequences. On the Pokec dataset, sampling-based NodeFormer with 100K sequence length outperforms the case with 10K
sequence length by a staggering 12\% accuracy. These results necessitate the need for long sequence training of graph transformers.


\subsection{Issues and Opportunities}
\label{subsec_issues}
Most existing graph transformer works \cite{Graphormer,GT,GraphGPS,Exphormer,Gophormer,SAT} are only limited on small graphs due to a lack of compatible systems tailored for the graph transformer model training with long sequences. They have three main issues when applied to long sequence training.

\noindent\textbf{I1: Attention Computation Bottleneck.} 
Graph transformers with standard (dense) attention treat the graph as fully-connected with the MHA mechanism calculating attention for all node pairs. Thus, it requires the computation complexity of the attention module to be quadratic on the number of nodes ($N^2$) in a graph, which limits the models' scalability to extremely long sequences. Currently, there is a breakthrough in standard attention optimization, i.e., FlashAttention \cite{FlashAttention}. FlashAttention accelerates the attention module by fusing the IO-bound GPU kernels like \texttt{Softmax} and \texttt{Dropout} within the attention computation. 
However, even with FlashAttention to train graph transformers with long sequences, e.g., sequence length of 512K, we still identify that the attention module dominates the overall training time. 

\begin{figure}[t]
    \centering
    \includegraphics[width=0.9\linewidth]{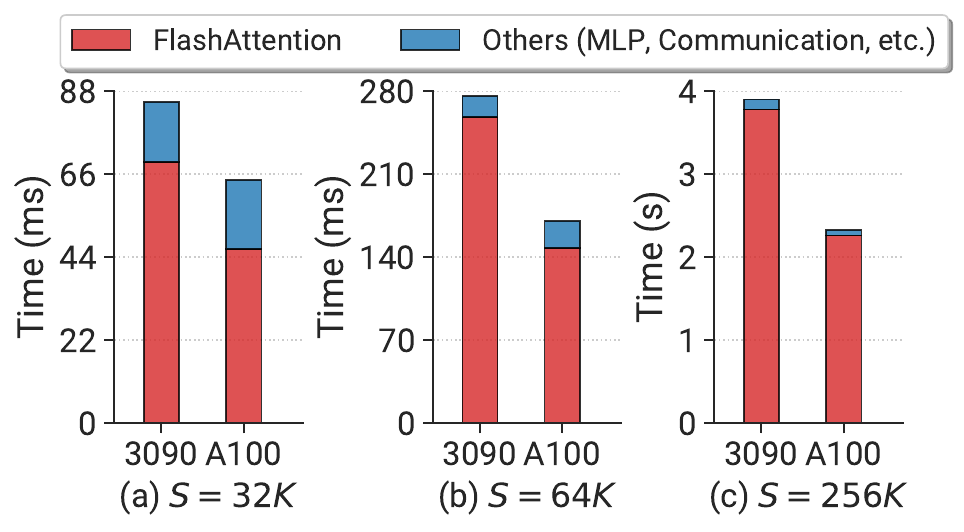}
    \caption{Training iteration time breakdown when training Graphormer on ogbn-products in different sequence lengths on two types of GPUs: RTX 3090 and A100.}
    \label{fig_motivation_flashattn}
\end{figure}

To show this, we conduct an experiment to record the iteration time breakdown when using FlashAttention, as illustrated in Figure \ref{fig_motivation_flashattn}. Current FlashAttention does not support the modified attention module like those augmented with bias encodings \cite{Graphormer,GT,Gophormer,NodeFormer}, so we disable the bias in this experiment to only examine the computation efficiency. We separate the computation time of FlashAttention from the comprehensive training iteration. We can obviously observe that no matter on longer or shorter sequences, attention computation still dominates over 80\% of training time, indicating a severe attention bottleneck. However, both the standard attention and FlashAttention fail to leverage one important characteristic of the graph, namely its topology structure, which we find profoundly impacts the effectiveness of system optimizations. 

\noindent\textbf{I2: Degraded Model Convergence and Limited Tasks.} Many efforts \cite{NodeFormer,GraphGPS,Performer,SpAtten} have been made to overcome the computation bottleneck of the attention module. Among them, \cite{SpAtten} prunes the attention module and leaves a major backbone to reduce the computation cost for LLMs. Some works \cite{Performer,BIGBIRD,Linformer,Reformer} propose sparse patterns on attention to scale linearly, but most of them are designed for natural language processing (NLP). They cannot be simply grafted to graph transformers since they fail to consider the inherent graph structure information when approximating attention, thus resulting in subpar model performance. Several graph transformers \cite{NAGphormer,Gophormer,Gapformer} harness neighbor sampling or graph pooling that only selects a subset of nodes to be trained at each iteration, without reducing the computation complexity. Nonetheless, all the above methods sacrifice model precision by dropping the connectivity information.

In the graph domain, efficient attention is not well studied. Few graph transformers like \cite{Exphormer} apply the graph structure to attend nodes and maintain graph-specific information. However, they limit the implementation to the GNN-encoding-based model architecture, e.g., GraphGPS~\cite{GraphGPS}, and highly rely on the message-passing scheme for excellent model performance. Other methods \cite{SGFormer,NodeFormer,DIFFormer} use self-defined adapted attention modules to achieve linear complexity. However, all those works are constrained to a single application task, failing to generalize to versatile graph tasks. Additionally, with GNN structure encodings or self-defined attention, the model can hardly be scaled to multiple workers. 
\noindent\textbf{I3: Lack of Specific System Optimizations.} As far as we know, currently there is no existing framework to optimize graph transformer training from the system level. FFN operations in MHA are dense in computation and regular in memory access. However, utilizing graphs on the attention module is sparse in computation and requires irregular and fine-grained memory access due to the skewed nature of graph structures, which inevitably becomes the performance barrier. 

Existing solutions \cite{Exphormer,GraphGPS,GT} directly apply graph topology in the attention computation while ignoring the pattern differences between graph transformer and standard Transformer-based models. To better illustrate, we experimentally examine the impact of irregular memory access by the topology-pattern attention in Table \ref{table_irregular_mm}. The topology-induced memory access latency is tremendous, reaching up to 33.2$\times$ slowdown than dense computation. To increase models' scalability, recent works \cite{sp-colossalai,sp-ds,sp-megatron,BlockwiseRingAttn} split the input sequences and distribute the computation across devices. However, this parallelism neglects various graph encoding modules and fails to distinguish input tokens in graph domain from tokens in NLP. Those differences invoke specialized and dedicated system designs for graph transformers towards more efficient memory optimizations and more aggressive parallelism.

\begin{table}[t]
    \centering
    \caption{Backward (BW) time of topology-pattern \& dense counterpart when training Graphormer on ogbn-products.}
    \resizebox{\linewidth}{!}{
        \begin{tabular}{@{}ccccc@{}}
\toprule
\textbf{Seq.   Length}                                                                                & $S$=64K                 & $S$=128K                & $S$=256K                 & $S$=512K                \\ \midrule
\multirow{2}{*}{\textbf{\begin{tabular}[c]{@{}c@{}}Topology-pattern\\      BW. Time/ms\end{tabular}}} & \multirow{2}{*}{116.99} & \multirow{2}{*}{234.28} & \multirow{2}{*}{499.289} & \multirow{2}{*}{963.91} \\
                                                                                                      &                         &                         &                          &                         \\
\textbf{Dense BW. Time/ms}                                                                            & 1.53                    & 3.78                    & 27.62                    & 29.01                   \\ \bottomrule
\end{tabular}
}
\label{table_irregular_mm}
\end{table}

%% file: 3_System_Design.tex
\section{\SysName Design}
\label{sec_primo}


We propose \SysName, an algorithm-system co-optimized system tailored for graph transformer training on large-scale graphs. It follows four design principles:

\begin{itemize}[leftmargin=*, itemsep=0pt, topsep=2pt]

\item \textit{Scalable}. \SysName can scale graph transformer training to extremely large graphs (solving \textbf{I3}). 

\item \textit{Efficient}. \SysName reduces over 90\% computation required by standard attention, overcoming the attention computation bottleneck (solving \textbf{I1}). 
\item \textit{Convergence-maintained}. \SysName maintains comparable model convergence to the graph transformer with standard attention, and successfully balances the trade-off between training efficiency and model quality (solving \textbf{I2}). 
\item \textit{Task-agnostic}. \SysName generalizes to various graph transformer models and graph learning tasks (graph-level and node-level) (solving \textbf{I2}).

\end{itemize}

\subsection{System Overview}

Motivated by all the observations in \S \ref{subsec_issues}, our key idea is to co-design an accuracy-maintained and compute-efficient attention module with a graph-parallelism-enabled system framework to support long sequence training. As shown in Figure \ref{fig_overview}, \SysName intelligently optimizes training across three levels from the top to bottom hierarchy: \textit{algorithm}, \textit{runtime} and \textit{kernel}. We propose a topology-induced and accurate attention algorithm in the algorithm level. We present a novel cluster-aware graph parallelism to scale the training in the runtime level. In the kernel level, we design a memory-optimized computation pattern specialized for clustered attention. Specifically, \SysName consists of three key modules:

\begin{figure}[t]
	\centering
	{\includegraphics[width=\linewidth]{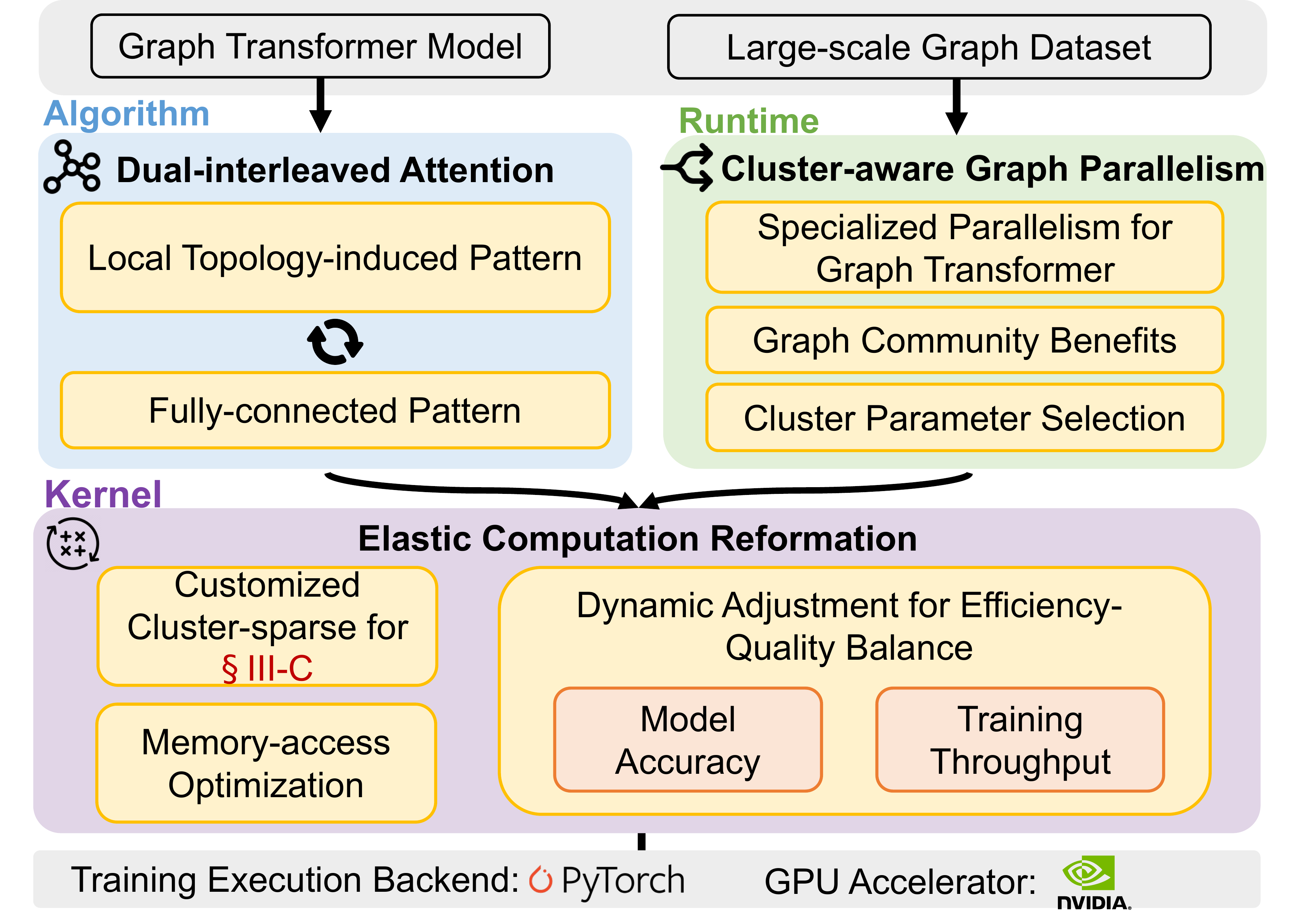} 
	\caption{Overview of \SysName architecture and workflow.}
	\label{fig_overview}}
\end{figure}


\begin{itemize}[leftmargin=*, itemsep=0pt, topsep=2pt]

\item \textit{Dual-interleaved Attention}: In the algorithm level, it integrates locally graph-induced topology into the attention computation pattern and periodically overlays it with the fully-connected information, which efficiently reduces the computation burden while maintaining the model's quality as much as possible. It is tailored for versatile graph transformer models with local-global interleaved attention.

\item \textit{Cluster-aware Graph Parallelism}: From the distributed runtime perspective, we design a cluster-aware graph parallelism tailored for graph transformers. It splits the graph tokens in sequences according to the clustering nature of graphs, thus computing attention with locality and facilitating the system scalability.

\item \textit{Elastic Computation Reformation}: It reformats the graph-induced pattern obtained at the runtime level into our customized and fine-grained cluster-sparse pattern at the underlying execution kernel level. It further improves the attention computation throughput by greatly alleviating irregular memory access. To balance the efficiency-quality trade-off, we build an \textit{Auto Tuner} to make an elastic transfer of cluster sparsity.

\end{itemize}

\subsection{Dual-interleaved Attention}
\label{subsec_attn}
Motivated by \textbf{I1} in \S \ref{subsec_issues}, \SysName explores the opportunity of integrating graph structure to reduce the substantial computation cost. For optimizations aiming at graph transformers, we design an interleaved attention to realize a local-global interleaved attention and ensure model convergence. 

\noindent\textbf{Local Topology-induced Pattern.} In NLP tasks, the tokens in a sequence represent words, while in graph transformers the tokens are nodes of the input graph. Besides, most graph transformers like \cite{Graphormer,NodeFormer} adopt the standard attention, which can be viewed as a fully-connected graph since all tokens attend to every other token to perform inner products, leading to quadratic complexity. Motivated by the sparse attention methods \cite{BIGBIRD,Performer}, we find the local topology-induced pattern that makes use of the underlying structure of the input graph is desirable to guide the pair-wise node interactions. Graphs innately own two desiderata for attention mechanism: (1) \textit{small pair-wise node interactions (large sparsity)}, and (2) \textit{data locality}. In addition, most sparse patterns in NLP are only approximations \cite{BIGBIRD} to their dense counterparts under specific contexts, while in our scenario the graph structure is real and valid, without the need of approximations. 
Thus, we compute attention by attending each node to its immediate neighbors in the graph, reducing the interacted node pairs. 

We formulate the local topology-induced pattern as below. To train on a graph $G=(V,E)$, we generate an input sequence $\boldsymbol{S} \in \mathbb{R}^{S \times d}$ comprised of graph tokens corresponding to a node set $\tilde{V}\in V$. $\tilde{V}$ can be equal to either the whole nodes $V$, e.g., in graph-level tasks, or a subset of $V$, e.g., in node-level tasks if the node number is too large. For each node set $\tilde{V}$, we construct a local attention graph $\tilde{G}=(\tilde{V},\tilde{E})$, where the edge set $\tilde{E}$ is also a subset of the original edge set $E$.
If there exists a global token in the model that attends to all nodes in the graph and is attended to by all nodes, we augment $\tilde{E}$ with the global token's edges. 
The general attention coefficient $\tilde{\boldsymbol{A}}_{i j}$ of graph transformers without graph encodings is computed in: $\tilde{\boldsymbol{A}}_{i j} = \text{softmax} \left( \frac{(h_i \boldsymbol{W_{\tilde{Q}}}) \cdot (h_j \boldsymbol{W_{\tilde{K}}})^{\top}}{\sqrt{d_{\tilde{\boldsymbol{K}}}}} \right)$. The updated node attribute $h'_i$ for each node $i$ is computed as the weighted sum of the features of its neighboring nodes from $\tilde{V}$: $h'_i = \sum_{j \in \mathcal{N}(i)} \tilde{\boldsymbol{A}}_{i j} (h_j \boldsymbol{W_{\tilde{V}}})$, where $\mathcal{N}(i)$ denotes the set of neighboring nodes of node $i$. Each neighbor's feature vector $h_j$ is weighted by the attention coefficient $\tilde{\boldsymbol{A}}_{i j}$, and these weighted features are summed to update the attribute of node $i$. By using our local topology-induced pattern $\tilde{G}$, the attention only computes coefficients of connected node pairs.




\noindent\textbf{Interleave Fully-connected Pattern.}
Implementing the attention computation via the graph structure can greatly reduce the computation cost. However, it sometimes slows the model accuracy and convergence, which can be shown by experiment results in Figure \ref{fig_attn_converge}. The local graph-induced attention slightly degrades the model convergence,
which is mainly because the topology-induced pattern restricts the attention mechanism from extracting the high-order neighboring information. 
Intuitively, larger sparsity induces more absence in the attention computation, and increases the model error. Building on this, we empirically interleave a fully-connected attention on the local graph-induced attention. 

To fill the performance gap between sparse attention and its dense counterpart, we need to figure out when to interleave the dense pattern. Motivated by the sparse attention theories in \cite{TransformerApproximability}, we conclude three critical conditions under which we use the topology-induced pattern on attention:

\begin{itemize}[topsep=2pt, leftmargin=*,itemsep=0pt]
\item\textbf{C1:} Every node in the sequence $\boldsymbol{S}$ always attends to itself.

\item\textbf{C2:} There exists a Hamiltonian path that directly connects all nodes $\tilde{V}$ in the sequence. 

\item\textbf{C3:} All nodes in the sequence should be able to attend to other nodes, either directly or indirectly after $L$ graph transformer attention layers.
\end{itemize}

The Hamiltonian path \cite{Hamiltonian} or traceable path is a path in a graph that visits each node exactly once. For each graph $\tilde{G}$ corresponding to the input sequence, the \textit{Dual-interleaved Attention} module searches it across the above three conditions. We use a heuristic approach Dirac's theorem \cite{Dirac} to do quick checks so the overhead is negligible in epoch time.
If it satisfies these conditions, 
we perform attention computation with the topology-induced sparse pattern. Otherwise, \SysName heuristically determines the current sparse pattern may introduce more errors 
and we utilize the fully-connected attention mechanism in this case to ensure model quality.

\noindent\textbf{Computation \& Memory Complexity.} 
The topology of many real-world graphs can be immensely sparse. For instance, the ogbn-arxiv graph has 169K nodes and 1.2M edges, resulting in a sparsity of $4.1\times10^{-5}$ (the proportion of nonzero elements in the whole adjacency matrix). 
As a result, the local topology-induced attention significantly reduces the computation and memory-access complexity from $O(N^2)$ to $O(\tilde{E})$. Though we interleave several fully-connected attention occasionally, the overall computation efficiency is still improved significantly.

\noindent\textbf{Model Convergence.}
Graph-centric attention \cite{Exphormer} and classical GNNs \cite{GraphSAGE,GAT} prove that sparse attention can maintain the model convergence comparable to its dense counterpart.  \cite{TransformerApproximability,SAN} propose sparse attention can obtain similar universality as dense attention under some assumptions. Borrowing it to \SysName, our \textit{Dual-interleaved Attention} can provide universal approximation properties that every continuous function $f$ can be approximated to any desired accuracy using a suitable sparse pattern under the three conditions, thus obtaining convergence similar to dense counterparts. 

\subsection{Cluster-aware Graph Parallelism}
\label{subsec_gp}
To better fit the topology-induced attention pattern and increase the system scalability, we introduce a graph transformer-specialized parallel training style: \textit{Cluster-aware Graph Parallelism}, which exploits the graph cluster characteristics to guide the distributed training.

\noindent \textbf{Utilization of Graph Cluster.} 
Graph cluster (community) \cite{Community1,SpectralCommunityDetect} is one essential characteristic of real-world graphs, referring to a subset of nodes within a graph that exhibit a higher degree of connectivity with each other compared to nodes in other parts of the graph. Although the graph structure-based attention in \S \ref{subsec_attn} greatly reduces the computation, this sparse and highly-skewed nature of graphs triggers substantial irregular memory access since edge connections are distributed in an uneven pattern, bringing extra overhead to training. Consequently, employing the graph cluster structure on GPUs is promising for graph transformer training improvement. There exist some approaches in traditional graph learning \cite{GraphPartition,SpectralCommunityDetect} to utilize graph cluster, but they are aimed for CPU processing with limited parallelization. \cite{GNNAdvisor} also exploits graph cluster but focuses on redundant data loading in GNN computing. 

Therefore, to explore the performance benefits of graph cluster on graph attention computing, we incorporate a lightweight node reordering to cluster nodes and improve the spatial locality during attention computation, without changing the connectivity correctness. The key insight is that the proximity of node IDs is more likely to be scheduled to the adjacency of computing units on GPUs where they get processed. In detail, we leverage METIS \cite{METIS}, a community-based graph reordering technique for great cluster locality and ease of integration with parallelism. Specifically, it uses multilevel recursive bipartitioning to divide and coarsen the graph while preserving the essential structure. 
We optimize the implementation of METIS for a lower cost: we capture the cluster information of graphs and map such locality from the upper level to the underlying GPU kernels, which also enables us to leverage the L1 \& L2 cache for refined cluster capturing (later discussed in \S \ref{subsec_block}). 

\noindent \textbf{Specialized Graph Parallelism.}
To increase the scalability of graph transformers, intuitively \SysName employs parallelism technologies to dispense the computation across devices. There have been extensive studies in sequence parallelism technologies \cite{sp-colossalai,sp-ds,sp-megatron} for LLMs to support efficient long sequence training. However, current parallelism methods for language models trigger two challenges when applied to graph transformers: (1) failing to leverage graph properties; (2) not supporting various graph encodings. In traditional language models, the input sequence encodes the context of a specific sentence. As such, training the language model requires tokens in the input sequence concatenated in a pre-defined order. In contrast, we observe that for graph learning tasks, \textit{there is no need for graph transformers to predict sequences (graph-level task) or tokens (node-level task) within a position-fixed context}, since they only rely on the graph topology to construct the structural encodings. A motivating example of parallelizing graphs with graph cluster is the graph-level task, where only the global token is critical for inferring the graph type and other node tokens can be arranged in any order. 

\begin{figure}[t]
	\centering
	{\includegraphics[width=\linewidth]{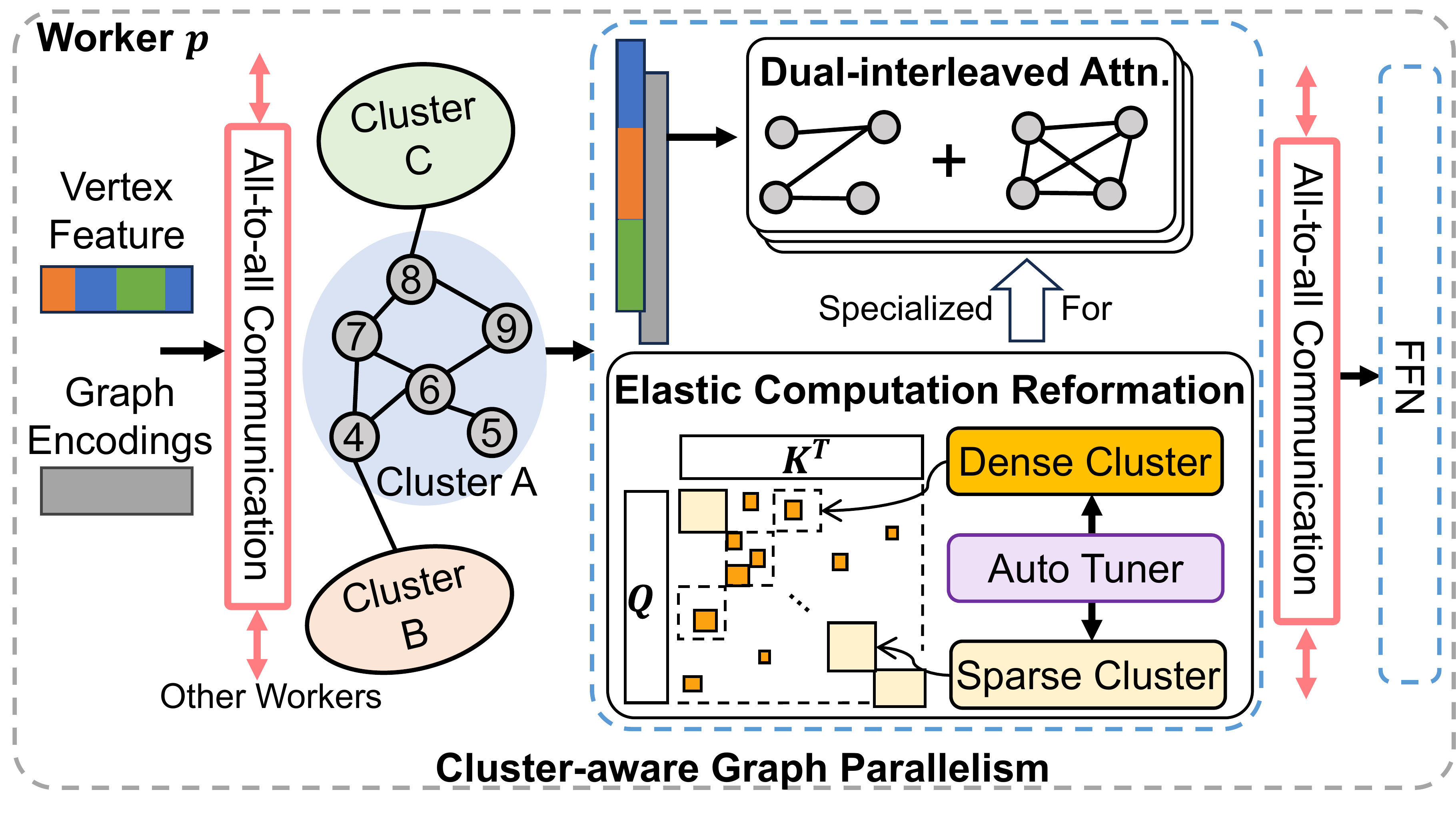} 
	\caption{Detailed training process on one worker with \SysName, which includes three key components.}
	\label{fig_detailed}}
\end{figure}

Based on this insight, we are the \textbf{\textit{first}} to design a \textit{Cluster-aware Graph Parallelism} \textbf{\textit{specialized for graph transformers}}, as shown in Figure \ref{fig_detailed}. Specifically, the raw input sequences $\boldsymbol{S}$ and graph encodings $\boldsymbol{B}$ are randomly partitioned across $P$ devices. Each local sub-sequence $\boldsymbol{S}_{sub}$ and sub-encodings are projected to local matrices: $\boldsymbol{Q}_{sub}, \boldsymbol{K}_{sub}, \boldsymbol{V}_{sub}, \boldsymbol{B}_{sub}\in \mathbb{R}^{\frac{S}{P} \times d}$, assuming they have the same dimensionality. Then in each graph transformer layer, all subspaces are combined together into complete matrices $\boldsymbol{Q},\boldsymbol{K},\boldsymbol{V}\text{, and }\boldsymbol{B}$ via the highly efficient all-to-all collective communication operation. All-to-all operation owns an advantage over other communication operations (e.g., all-gather and reduce-scatter) in terms of much smaller communication volume and overall better scalability, which is also proved in \cite{sp-ds}. All-to-all gathers matrices in sequence dimension and splits in the head, resulting in $\boldsymbol{Q},\boldsymbol{K},\boldsymbol{V}\text{, and }\boldsymbol{B}\in \mathbb{R}^{S\times \frac{d}{P}}$. Now that since matrices are complete in the sequence dimension, \SysName reorganizes the layout according to the clustering nature of graphs discussed before. Then the \textit{Dual-interleaved Attention} conducts attention computation in the clustered layout, exemplified as blue, orange and green rectangles in Figure \ref{fig_detailed}. After attention computation, another all-to-all transforms the output tensor back to subspace $\boldsymbol{S}'_{sub}$ for subsequent operators such as FFN and layer normalization in the graph transformer layer.

\noindent\textbf{Communication Complexity.} Thanks to all-to-all, \textit{Cluster-aware Graph Parallelism} has low communication volume and scales exactly well with more servers. Given hidden size $d$, sequence length $S$, and parallelism degree $P$, 
\SysName performs all-to-all with a total message size $3Sd$ before the attention computation,
and another all-to-all for attention output with size $Sd$. Therefore, \SysName performs two all-to-alls with communication volume per GPU of $4Sd/P$ and communication complexity of $O(S/P)$, while other operations like all-gather have communication complexity of $O(S)$. Thus, \SysName has better scalability with increasing parallelism degree on extremely long sequences.

In summary, compared with sequence parallelism methods for LLMs, our \textit{Cluster-aware Graph Parallelism} favors the graph transformer architecture in several aspects. First, all-to-all gathers in sequence dimension, leading to exactly integrated graph topology, which the topology-induced sparse pattern in \S \ref{subsec_attn} can be perfectly applied to. Second, the graph encodings $\boldsymbol{B}$ share the same sparse layout as attention mapping so the parallelism of graph transformers only brings a trivial memory footprint and communication overhead,  thus facilitating model scalability and ensuring memory efficiency. 


\begin{figure}[t]
	\centering
	{\includegraphics[width=\linewidth]{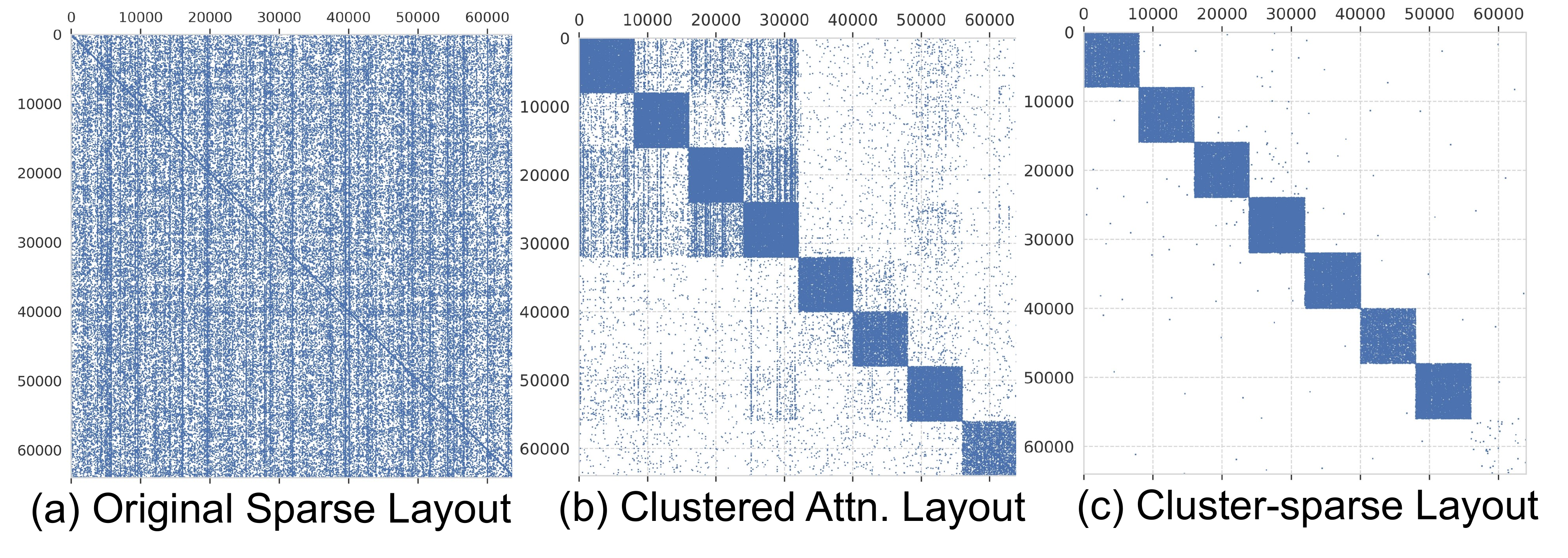} 
	\caption{Three attention layouts after \textit{Dual-interleaved Attention}, \textit{Cluster-aware Graph Parallelism} and \textit{Elastic Computation Reformation} respectively. (c) is obtained by compacting elements to adjacent neighbors inside clusters.}
	\label{fig_sparse-layout}}
\end{figure}

\subsection{Elastic Computation Reformation}
\label{subsec_block}
\noindent \textbf{Cluster Sparsity.}
The topology-induced attention pattern can significantly reduce the computation cost, but also leading to substantial irregular memory access due to the highly skewed nature of graphs. 
Figure \ref{fig_sparse-layout}(b) gives an example with the cluster dimensionality of 8 and sequence length of 64K. We observe that only the diagonal clusters in the clustered adjacency matrix appear in dense patterns most and show lower sparsity, which can benefit from locality since nodes in each cluster are close to each other. On the other hand, other clusters appear highly sparse patterns and more irregular shapes (denoted as \textit{sparse cluster}). Accessing the embeddings of computation like aggregation in this pattern requires a large number of atomic operations. Consequently, those remaining irregular clusters still engender heavy overhead.
To exemplify, training Graphormer on ogbn-arxiv ($S$=64K) in Figure \ref{fig_sparse-layout}(b) takes 375 ms per epoch, while its dense counterpart only takes 81ms.

To further reduce the memory access latency, we propose a memory-efficient \textit{Elastic Computation Reformation} which introduces the cluster sparsity. Motivated by the block-sparse pattern in \cite{BIGBIRD,BlocksparseRNN,BlockBert}, we reformat the clustered layout in Figure \ref{fig_sparse-layout}(b) to a fine-grained cluster-level fashion in Figure \ref{fig_sparse-layout}(c). Specifically, as shown in Figure \ref{fig_detailed}, for each sparse cluster, \SysName transfers the scattered edges inside it to multiple substructures of compact and adjacent edge connections, which is denoted as \textit{sub-blocks}. The transferred \textit{dense cluster} can have multiple randomly scattered sub-blocks, the number of which is decided by the number of real edges in the cluster and the dimension of sub-block $d_{b}$. Note that there will be totally $k^2$ clusters for the whole layout with the cluster dimensionality of $k$. Figure \ref{fig_sparse-layout}(c) depicts the cluster-sparse layout with $k$=8, in which most sparse clusters are transferred to dense ones with tight sub-blocks.
This cluster sparsity offsets the downside of irregular memory access incurred by the topology-induced pattern.


\noindent \textbf{Transfer Strategy.}
Applying static cluster-sparse transferring will result in model quality degradation since the cluster sparsity changes the original graph structure by modifying edges. 
Some performance-related information (e.g., convergence) for model training is only available at runtime. Without considering the runtime information, the system will suffer from an inferior model accuracy or inefficient memory access. 

Thus, \SysName designs the following two strategies:

\begin{itemize}[topsep=2pt, leftmargin=*,itemsep=0pt]
\item \textit{Indolent Transferring.} \SysName only transfers clusters that are extremely sparse and irregular. Although such an inactive way may miss some optimization opportunities, it can refrain from model quality decline and be more portable. Concretely, \SysName only transfers sparse clusters whose sparsity value $\beta_C$ is smaller than that of the current whole graph $\beta_G$. Note that the sparsity value refers to the proportion of \textbf{nonzero} elements in the whole adjacency matrix.

\item \textit{Elastic Transferring.} We dynamically adjust the amount of transferred dense clusters along the training. First, we set a threshold value $\beta_{thre}$ for controlling cluster-sparse transfer. If the sparsity of a cluster $\beta_C$ is smaller than the threshold $\beta_{thre}$, \SysName transfers it to a dense cluster. To decide the value of $\beta_{thre}$ in each training epoch, we design an \textit{Auto Tuner} in the next part for modeling $\beta_{thre}$.

\end{itemize}

\begin{figure}[t]
	\centering
	{\includegraphics[width=\linewidth]{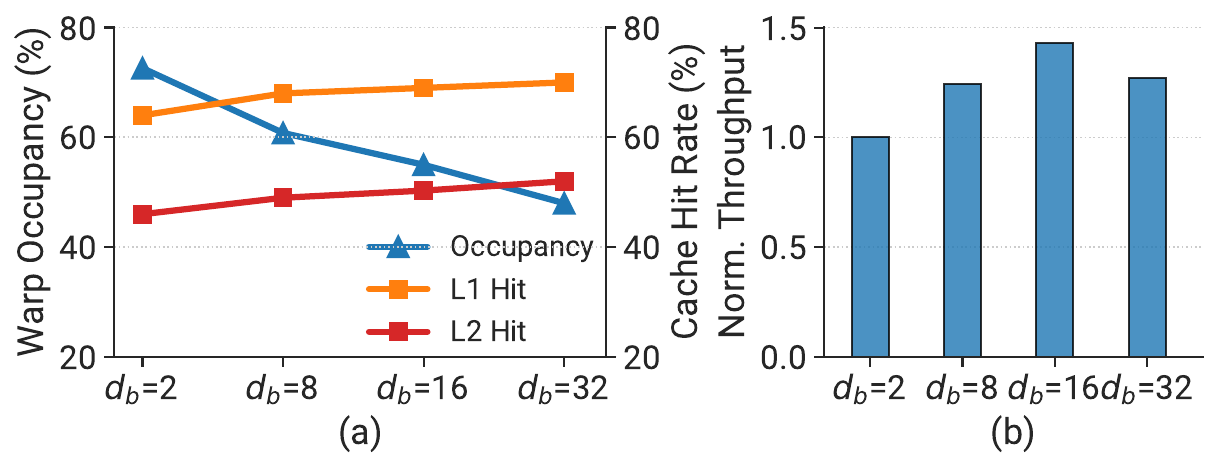}
	\caption{(a) Profiled hardware statistics of GPU when computing in different sub-block size $d_b$. The ideal $d_b$ considers both load balance and cache hit rate. (b) Computation throughput of the indexing kernel in different $d_b$ normalized on that of $d_b$=2.}
	\label{fig_blocks_hint}}
\end{figure}

\noindent \textbf{Hyperparameter Modeling.} The hyperparameters can be tuned to accommodate various graph patterns. We design an \textit{Auto Tuner} to dynamically select the hyperparameters $k$, $d_b$ and $\beta_{thre}$, and formulate the analytical model. 

\textit{(1) Cluster dimensionality $k$, sub-block dimension $d_b$.} We can leverage GPU L1 and L2 caches to improve the memory access locality of sub-block computation.
In this way, smaller sub-blocks can enjoy the data locality benefit from the L1 cache while larger clusters can enjoy the locality from the larger L2 cache. Specifically, we determine $k$ as: $k=\lfloor \sqrt{\frac{Q_{\text{L2}}}{i d}} \rfloor,i\in \mathbb{N}$, where $Q_{\text{L2}}$ is the L2 cache size and $d$ is model hidden dimension. To determine the sub-block dimension $d_b$, we profile the computation throughput and some hardware statistics of the indexing kernel w.r.t. different $d_b$ values. Figure \ref{fig_blocks_hint}(a) shows the workload balance in GPU computing unit downgrades (the lower warp occupancy, the worse balance) as $d_b$ increases, while both L1 \& L2 cache hit rates increase. Thus, there exists a trade-off between these two metrics in deciding $d_b$. Moreover, Figure \ref{fig_blocks_hint}(b) also demonstrates the values of obtaining the optimal computation throughput lie in the middle range. Both cases suggest the middle value is the ideal choice. For example, for RTX 3090 GPU and model hidden dimension of 64, we fit $k$=8 and $d_b$=16. These two values can also be selected by users.

\textit{(2) Transfer threshold $\beta_{thre}$.} Motivated by \cite{Sylvie}, to estimate the convergence, \textit{Auto Tuner} tracks a running average loss $F_t=0.9 F_{t-1}+0.1 \mathcal{L}_t$, where $\mathcal{L}_t$ is the loss at epoch $t$. Considering the training throughput, a Loss Descent Rate (LDR) is defined as $LDR_t=\frac{F_t - F_{t-1}}{et_{t}}$, where $et_t$ is the $t$-th epoch training time. At the beginning $\beta_{thre,0}$ is initialized as $\beta_{G}$ from the set $\{0,\beta_{G},1.5\beta_{G},5\beta_{G},7\beta_{G},10\beta_{G},1\}$, which is developed by profiling different datasets. When $LDR_t \geq LDR_{t-\delta}$ for some $\delta \in \mathbb{N}$ (here we use $\delta=10$) which specifies the range of epochs for $LDR$ comparisons, \SysName heuristically determines the current $\beta_{thre}$ suffices to reduce the loss. Then \textit{Auto Tuner} increases $\beta_{thre}$ to the next value in the set to gain higher speed. On the other hand, $LDR_t < LDR_{t-\delta}$ in $\delta$ epochs denotes the training is about to converge or too many errors are introduced by quantization. In this case, \textit{Auto Tuner} reduces $\beta_{thre}$ to the previous value from the set to enable more stable and accurate training. 




\begin{table}[t]
    \centering
    \caption{Detailed information of datasets in evaluation.}
    \resizebox{\linewidth}{!}{
        \begin{tabular}{@{}ccccc@{}}
\toprule
\multicolumn{5}{c}{\textbf{Node-level}}                                                                                                                                                                          \\ \midrule
\textbf{Datasets} & \textbf{\# Nodes} & \textbf{\# Edges}                                                         & \textbf{\# Feats}                                                      & \textbf{Task}      \\ \midrule
Amazon \cite{amazon}            & 1,598,960            & 132,169,734                                                               & 200                                                                    & 107-class Classif. \\
ogbn-arxiv \cite{OGB}       & 169,343              & 1,166,243                                                                 & 128                                                                    & 40-class Classif.  \\
ogbn-products \cite{OGB}    & 2,449,029            & 61,859,140                                                                & 100                                                                    & 47-class Classif.  \\
ogbn-papers100M \cite{OGB}  & 111,059,956          & 1,615,685,872                                                             & 128                                                                    & Binary Classif.    \\ \midrule
\multicolumn{5}{c}{\textbf{Graph-level}}                                                                                                                                                                           \\ \midrule
\textbf{Datasets} & \textbf{\# Graphs}   & \textbf{\begin{tabular}[c]{@{}c@{}}Avg. \\ \#   Nodes\end{tabular}} & \textbf{\begin{tabular}[c]{@{}c@{}}Avg. \\ \#  Edges\end{tabular}} & \textbf{Task}      \\ \midrule
ZINC \cite{zinc}             & 12,000               & 23.2                                                                      & 24.9                                                                   & Regression         \\
ogbg-molpcba \cite{OGB}     & 437,929              & 26.0                                                                        & 28.1                                                                   & 128-task Classif.  \\
MalNet \cite{Malnet}           & 10,833               & 15,378                                                                    & 35,167                                                                 & 5-class Classif.   \\ \bottomrule
\end{tabular}
            }
    \label{table_datasets}
\end{table}

\begin{table}[t]
    \centering
    \caption{Detailed information of graph transformer models.}
    \resizebox{0.9\linewidth}{!}{
        \begin{tabular}{@{}cccc@{}}
\toprule
\textbf{Models}                        & \textbf{\# Layers} & \textbf{Hidden Dim.} & \textbf{\# Head} \\ \midrule
Graphormer$_{slim}$   (GPH$_{slim}$)   & 4                  & 64                   & 8                \\
Graphormer$_{large}$   (GPH$_{large}$) & 12                 & 768                  & 32               \\
GT                                     & 4                  & 128                  & 8                \\ \bottomrule
\end{tabular}
            }
    \label{table_model}
\end{table}

\begin{table*}[t]
    \centering
        \caption{Detailed comparison of training speed and test accuracy of methods when training on one 3090 GPU server. OOM means the method runs out of memory. \SysName always outperforms \FlashGT in throughput and accuracy on all the models and datasets. \RawGT with full attention runs out of memory in all cases.}
    \resizebox{\linewidth}{!}{
      \begin{tabular}{@{}cccccccccccc@{}}
\toprule
\multirow{2}{*}{\textbf{Model}}         & \multirow{2}{*}{\textbf{Method}} & \multicolumn{2}{c}{\textbf{MalNet}}                    & \multicolumn{2}{c}{\textbf{ogbn-papers100m}}          & \multicolumn{2}{c}{\textbf{ogbn-products}}           & \multicolumn{2}{c}{\textbf{ogbn-arxiv}}             & \multicolumn{2}{c}{\textbf{Amazon}}                  \\ \cmidrule(l){3-4} \cmidrule(l){5-6} \cmidrule(l){7-8} \cmidrule(l){9-10} \cmidrule(l){11-12} 
                                        &                                  & \textbf{$t_{epoch}$(s)}       & \textbf{Test Acc.(\%)} & \textbf{$t_{epoch}$(s)}      & \textbf{Test Acc.(\%)} & \textbf{$t_{epoch}$(s)}     & \textbf{Test Acc.(\%)} & \textbf{$t_{epoch}$(s)}    & \textbf{Test Acc.(\%)} & \textbf{$t_{epoch}$/s}      & \textbf{Test Acc.(\%)} \\ \midrule
\multirow{3}{*}{\textbf{GPH$_{Slim}$}}  & GP-Raw                           & OOM                           & -                      & OOM                          & -                      & OOM                         & -                      & OOM                        & -                      & OOM                         & -                      \\
                                        & GP-Flash                         & 2158.37                       & 90.87                  & 1201.13                      & 90.11                  & 27.69                       & 66.39                  & 0.44                       & 48.25                  & 17.31                       & 63.51                  \\
                                        & \SysName                         & 195.54(\textbf{11.0$\times$}) & 92.71                  & 19.15(\textbf{62.7$\times$}) & 96.82                  & 0.54(\textbf{50.8$\times$}) & 66.75                  & 0.11(\textbf{3.9$\times$}) & 53.81                  & 1.00(\textbf{17.5$\times$}) & 73.10                  \\ \midrule
\multirow{3}{*}{\textbf{GPH$_{Large}$}} & GP-Raw                           & \multicolumn{2}{c}{\multirow{3}{*}{OOM}}               & OOM                          & OOM                    & OOM                         & -                      & OOM                        & -                      & OOM                         & -                      \\
                                        & GP-Flash                         & \multicolumn{2}{c}{}                                   & 2512.88                      & 96.93                  & 56.51                       & 44.48                  & 3.46                       & 22.11                  & 36.83                       & 73.34                  \\
                                        & \SysName                         & \multicolumn{2}{c}{}                                   & 654.72(\textbf{3.8$\times$}) & 98.60                  & 16.10(\textbf{3.5$\times$}) & 63.06                  & 1.16(\textbf{3.0$\times$}) & 42.38                  & 11.07(\textbf{3.3$\times$}) & 73.75                  \\ \midrule
\multirow{3}{*}{\textbf{GT}}            & GP-Raw                           & OOM                           & -                      & OOM                          & -                      & OOM                         & -                      & OOM                        & -                      & OOM                         & -                      \\
                                        & GP-Flash                         & 1426.24                       & 74.54                  & 1235.02                      & 88.86                  & 28.80                       & 66.20                  & 0.50                       & 53.98                  & 8.88                        & 69.07                  \\
                                        & \SysName                         & 242.58(\textbf{5.9$\times$})  & 90.13                  & 26.33(\textbf{46.9$\times$}) & 89.60                  & 0.79(\textbf{36.3$\times$}) & 82.11                  & 0.09(\textbf{5.3$\times$}) & 56.72                  & 0.76(\textbf{11.7$\times$}) & 72.98                  \\ \bottomrule
\end{tabular}
            }
    \label{table_end2end}
\end{table*}

\begin{table}[t]
    \centering
    \caption{Training time per epoch of trianing GPH$_{Slim}$ on one A100 server. \SysName can still improve training efficiency compared with \FlashGT.}
    \resizebox{\linewidth}{!}{
      \begin{tabular}{@{}cccccc@{}}
\toprule
\multirow{2}{*}{\textbf{Model}}        & \multirow{2}{*}{\textbf{Method}} & \textbf{MalNet}              & \textbf{ogbn-papers100m}     & \textbf{ogbn-products}     & \textbf{Amazon}            \\ \cmidrule(l){3-6} 
                                       &                                  & \textbf{$t_{epoch}$(s)}      & \textbf{$t_{epoch}$(s)}      & \textbf{$t_{epoch}$(s)}    & \textbf{$t_{epoch}$(s)}    \\ \midrule
\multirow{2}{*}{\textbf{GPH$_{Slim}$}} & GP-Flash                         & 668.23                       & 492.79                       & 5.34                       & 3.43                       \\
                                       & \SysName                         & 160.61(\textbf{4.2$\times$}) & 244.07(\textbf{2.1$\times$}) & 2.86(\textbf{1.9$\times$}) & 1.69(\textbf{2.0$\times$}) \\ \bottomrule
\end{tabular}
            }
    \label{table_speed_a100}
\end{table}

%% file: 4_Evaluation.tex
\section{Evaluation}
\label{eval}



We implement \SysName atop PyTorch 2.1 \cite{PyTorch}. We study the performance of \SysName on versatile datasets and graph learning tasks in the following aspects: (1) Efficiency (\S \ref{exp_throughput}), (2) Convergence (\S \ref{exp_convergence}), (3) Scalability (\S \ref{exp_scale}), and (4) micro-benchmarks and ablation studies to examine the impact of each technique and hyperparameter (\S~\ref{exp_microbench}).

\noindent\textbf{Datasets and Models.} 
We evaluate \SysName on versatile real-world graph datasets with multiple scales. The detailed information is shown in Table \ref{table_datasets}, including both node-level and graph-level tasks. The MalNet dataset is constructed from all categories of the full datasets.  
We use three classical graph transformer models commonly adopted for evaluation, including Graphormer$_{Slim}$ (GPH$_{Slim}$) \cite{Graphormer}, Graphormer$_{Large}$ (GPH$_{Large}$) \cite{Graphormer}, and GT \cite{GT}. Note that \SysName can also be applied to other graph transformer models. As shown in Table \ref{table_model}, we follow the hyperparameter configurations reported in their original papers as closely as possible. 

\noindent\textbf{Baselines.} 
All models cannot be directly trained on selected large graphs. Due to the lack of existing graph transformer systems, we meticulously replicate each model with simple graph parallelism following its original implementation as the vanilla version, denoted as \RawGT. On the basis of this, we have also developed other variants incorporating FlashAttention~\cite{FlashAttention} denoted as \FlashGT, and topology-induced sparse attention denoted as \SparseGT.

\noindent \textbf{Testbed.}
Our experiments are performed on two GPU servers. \one 3 GPU servers each with 8 NVIDIA RTX 3090 GPUs (24GB). Intra-server connections (CPU-GPU, GPU-GPU) are based on PCIe 4.0x16 lanes and inter-server connections are via 1Gbps Ethernet. \two 2 servers each with 8 A100 GPUs (80GB) with NVLink and 200Gbps InfiniBand.

\subsection{End-to-end Training Throughput} 
\label{exp_throughput}
We compare the end-to-end training time per epoch and test accuracy of \SysName with all baselines on one server, as shown in Table \ref{table_end2end}. The sequence length is 256K for GPH$_{Slim}$ and GT, and 32K for GPH$_{Large}$. When training on ogbn-arxiv, we set the sequence length to 64K for GPH$_{Slim}$ and GT. The speedup in the bracket is the relative throughput of each method on the basis of \FlashGT. In each training task, we treat the first 10 epochs as the warmup stage and only record statistics afterward. 
\SysName substantially outperforms \FlashGT by 3.3$\sim$62.7$\times$. This is mainly because \SysName significantly reduces the computation complexity of the attention module. Additionally, \RawGT runs out of memory (OOM) on all datasets under the current sequence lengths due to its $O(N^2)$ memory complexity of the attention module. For instance, \RawGT requires over 200GB memory to store the attention score, i.e., $QK^{\top}$, of only one attention head for the ogbn-products dataset. We also conduct evaluations of GPH$_{Large}$ on one A100 server as shown in Table \ref{table_speed_a100}. \SysName still shows impressive acceleration and outperforms \FlashGT up to 4.2$\times$ on such frontier equipment. In summary, \SysName realizes efficient training of graph transformers with a marvelous improvement.

The speedup difference is mainly related to the input graph topology and model structure under long sequences. If the input graph is very sparse (up to 99\% sparsity), \textit{Dual-interleaved Attention} first boosts attention by a large margin. If an obvious clustering pattern exists in the graph, then attention can be further accelerated by 2$\sim$3$\times$ with the other two modules. 
Since we mainly improve attention computation, the more proportion it accounts in total model computation, the higher speedup on epoch time. For instance, in Table \ref{table_end2end} GPH$_{Slim}$ achieves notable speedup on papers100M owing to the above. Ogbn-arxiv shows a smaller speedup on all models since it has poorer sparsity and cluster property. 


\begin{table}[t]
    \centering
    \caption{Training throughput and test accuracy of methods. The accuracy of \FlashGT decreases because of BF16.}
    \resizebox{\linewidth}{!}{
      \begin{tabular}{@{}ccccc@{}}
\toprule
                                     & \multicolumn{1}{l}{} & \textbf{\FlashGT} & \textbf{\SysName-BF16} & \textbf{\SysName-FP32} \\ \midrule
\multirow{2}{*}{\textbf{ogbn-arxiv}} & $t_{epoch}$(s)        & 0.44              & \textbf{0.08}                   & 0.11                   \\
                                     & Test   Acc.(\%)       & 48.25             & 48.29                  & \textbf{53.81}                  \\ \midrule
\multirow{2}{*}{\textbf{Amazon}}     & $t_{epoch}$(s)        & 17.31             & \textbf{0.60}                   & 1.00                   \\
                                     & Test   Acc.(\%)       & 63.51             & 63.58                  & \textbf{73.10}                  \\ \bottomrule
\end{tabular}
            }
    \label{table_torchgtbf16}
    \vspace{-2pt}
\end{table}

\begin{figure}[t]
    \centering
    \includegraphics[width=\linewidth]{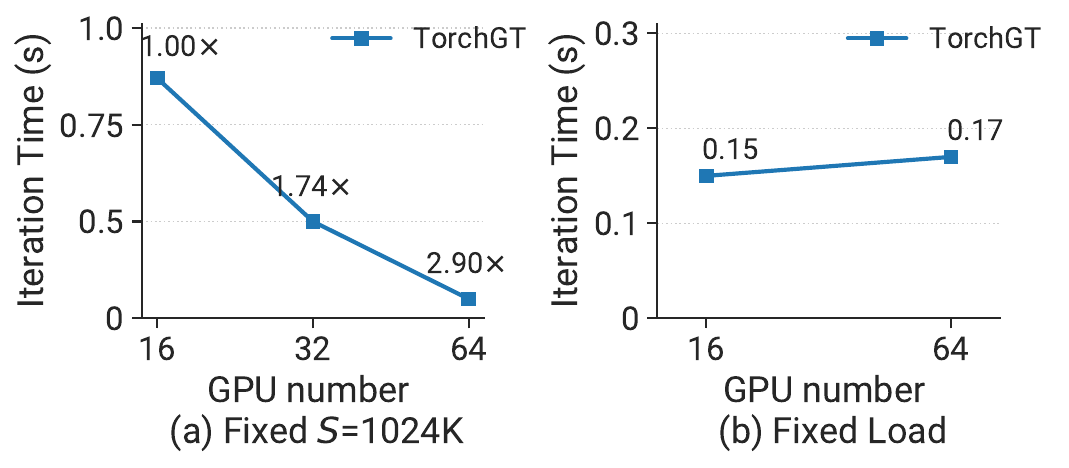}
    \caption{Scalability results of \SysName in training GPH$_{Slim}$ on ogbn-products on many A100 servers. (a) With fixed sequence length, throughput reduces almost linearly. (b) With fixed computational load per GPU, throughput remains well.}
    \label{fig_scalability2}
\end{figure}

\begin{figure*}[t]
    \centering
    \includegraphics[width=\textwidth]{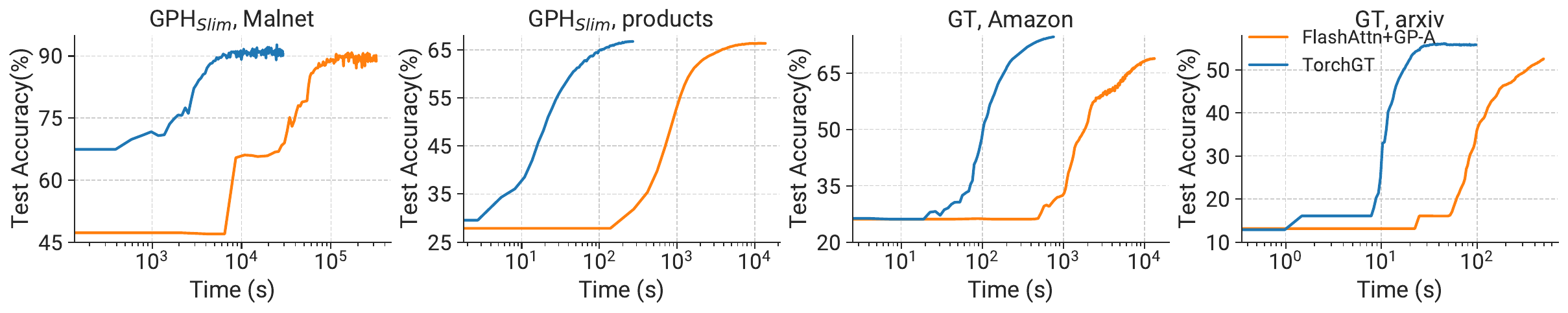}
    \caption{The convergence curve comparisons of \SysName and \FlashGT on different models and datasets.}
    \label{fig_end2end_converge}
    \vspace{-5pt}
\end{figure*}

\subsection{Model Convergence}
\label{exp_convergence}
We examine the model accuracy and convergence curves of \SysName on various models in Table \ref{table_end2end} and Figure \ref{fig_end2end_converge}. Table~\ref{table_end2end} summarizes the test accuracy achieved by all systems on three models. On large-scale datasets, \SysName gives higher model accuracy while \RawGT runs out of memory. \FlashGT harms the model accuracy in some datasets, e.g., Malnet and Amazon since FlashAttention only supports FP16/BF16 precision~\cite{FlashAttention} in computing which may downgrade model convergence compared to FP32 precision. In contrast, \SysName supports FP32 precision without compromising model accuracy. To better validate this, we compare the training throughput and test accuracy of \FlashGT and \SysName-BF16 in Table \ref{table_torchgtbf16}. On BF16, \SysName obtains similar accuracy with FlashAttention, indicating the accuracy drop of FlashAttention is mainly caused by reduced precision. \SysName even achieves higher speedup with BF16, but we choose to use FP32 since it gives higher accuracy with notable speedup.
From Figure \ref{fig_end2end_converge}, we also see that \FlashGT converges to low accuracy and lead to far slower convergence than our system, verifying it preserves model quality well.



\begin{figure}[t]
    \centering
    \includegraphics[width=\linewidth]{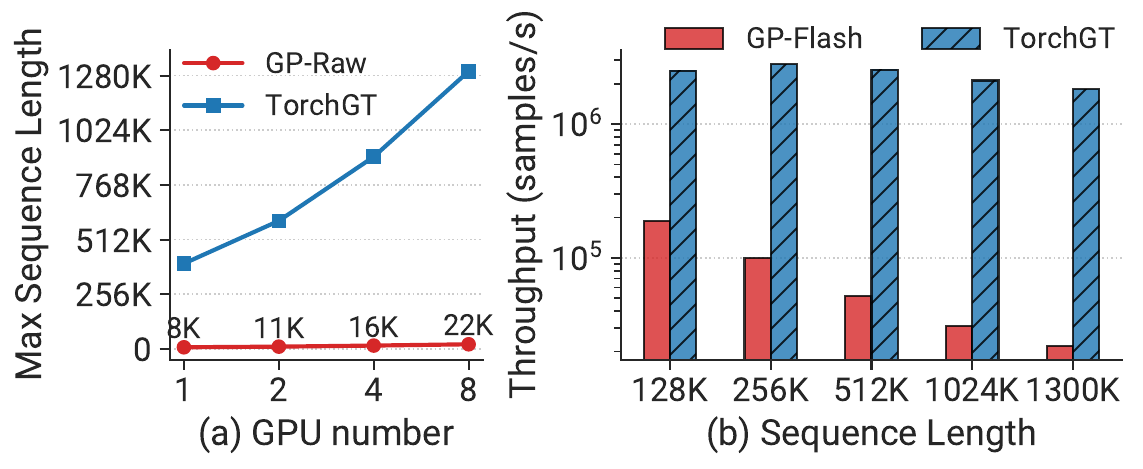}
    \caption{Scalability experiments of training GPH$_{Slim}$ on ogbgn-products. (a) The supported maximum sequence length w.r.t. GPU number. (b) Training throughput w.r.t. sequence length. In both cases \SysName shows greater scalability than others.}
    \label{fig_scalability}
\end{figure}

\subsection{System Scalability}
\label{exp_scale}


\noindent\textbf{Training Throughput on Multiple Servers.} First, we evaluate the training throughput of \SysName on multiple A100 servers to validate it also scales out well with more servers. As shown in Figure \ref{fig_scalability2}, we conduct two sets of scalability evaluations on up to 8 servers(each with 8 A100 GPUs) with extremely long sequences and record the sequence training time on the ogbn-products dataset. In Figure \ref{fig_scalability2}(a), we fix the sequence length to 1024K and increase the server number. We can see \SysName still obtains notable speedup when scaling to more servers. Especially, when the GPU count is doubled, the training throughput correspondingly increases by almost 1.7$\times$, indicating a certain degree of scalability. In Figure \ref{fig_scalability2}(b), we fix the computational load per GPU when increasing the sequence length from 256K to 512K. Note that when doubling the sequence length, we need 4$\times$ GPUs than before to keep the same computational load per GPU(attention calculation is proportional to $S^2/P$). In this case, \SysName achieves approximately the same throughput on each GPU as before, also verifying good scalability.

\noindent\textbf{Sequence Length w.r.t. Number of GPUs.} We examine the maximum sequence length of GPH$_{Slim}$ that can be trained on 1$\sim$8 GPUs with \SysName in Figure~\ref{fig_scalability}(a). Note that \RawGT employs standard full attention. We can see the maximum sequence length of \SysName can reach up to 1.3M on 8 GPUs. It also enables the sequence length of 400K with only 1 GPU, substantially 50$\times$ larger than that of \RawGT. Moreover, the sequence length of \SysName almost scales linearly w.r.t. the number of GPUs, while the maximum sequence length \RawGT can support nearly remains unchanged with the growth of GPU numbers. With 8 GPUs, \SysName supports 1.3M in length while \RawGT only supports 22K in length.

\noindent\textbf{Throughput w.r.t. Sequence Length.} We further compare the training throughput of \SysName and \FlashGT{} under sequence lengths varying from 128K to 1300K in Figure \ref{fig_scalability}(b). We fix the number of GPUs to 8 and report the throughput as samples per second. Figure~\ref{fig_scalability}(b) shows that the training throughput of \FlashGT{} sharply decreases from $1.9\times 10^5$ samples/s to $2.2\times 10^4$ samples/s when the sequence length increases. The speed degradation of \FlashGT mainly comes from the computation bottleneck of FlashAttention with $O(N^2)$ complexity. In contrast, \SysName maintains the training throughput at around $2.5\times 10^6$ samples/s by significantly reducing the attention computation costs (in \S \ref{subsec_attn}).

\subsection{Micro-benchmarks} 
\label{exp_microbench}
We explore the effects of each component in \SysName via ablation studies and perform sensitivity analysis of the introduced hyperparameters on one 3090 GPU server.

\noindent\textbf{Impact of \textit{Dual-interleaved Attention}.} After employing the topology-induced attention and interleaving mode introduced in \S \ref{subsec_attn}, we investigate their effect on model quality. 

\begin{figure}[t]
    \centering
    \includegraphics[width=\linewidth]{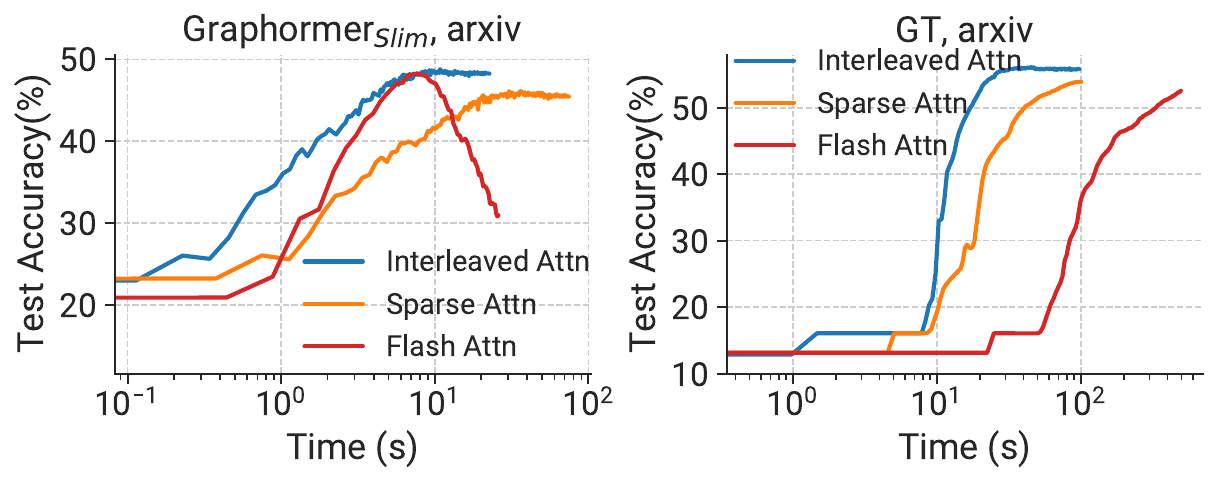}
    \caption{Convergence comparisons of different attentions: our interleaved attention, FlashAttention and sparse attention.}
    \label{fig_attn_converge}
\end{figure}

\textit{(1) On large-scale graphs.} We measure the convergence curves of GPH$_{Slim}$ and GT on ogbn-arxiv, and compare the convergence of interleaved attention with that of FlashAttention and the sparse variant in Figure \ref{fig_attn_converge}. The model with interleaved attention shows faster convergence than the other two and finally converges to higher accuracy, verifying that the interleaved attention improves computation efficiency while displaying great convergence.

\begin{figure}[t]
    \centering
    \includegraphics[width=\linewidth]{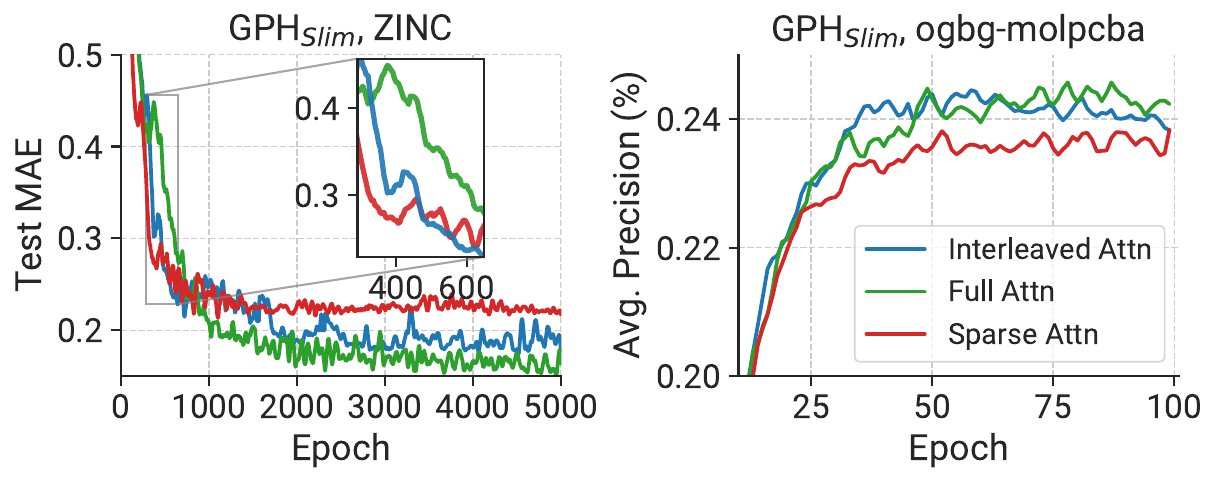}
    \caption{Convergence curves of different attentions: our interleaved attention, full and sparse attention on small graphs.}
    \label{fig_attn_converge2}
\end{figure}

\textit{(2) On small graphs.} Since the raw graph transformer models fail to be trained on large graphs, we further evaluate the convergence of interleaved attention on small graphs in Figure \ref{fig_attn_converge2}. Sparse attention shows the worst convergence rate while full attention has the best. The model with interleaved attention converges to nearly the same as the model with full attention and obviously outperforms the sparse variant in both convergence speed and final test score. 

\noindent\textbf{Impact of \textit{Elastic Computation Reformation}.}
We particularly examine the impact of the \textit{Elastic Computation Reformation} module in \SysName, FlashAttention, and sparse attention w.r.t. the sequence length and the model hidden dimension on one GPU. Note that we implement the sparse variant with the pure topology-induced attention pattern.

\begin{figure}[t]
    \centering
    \includegraphics[width=\linewidth]{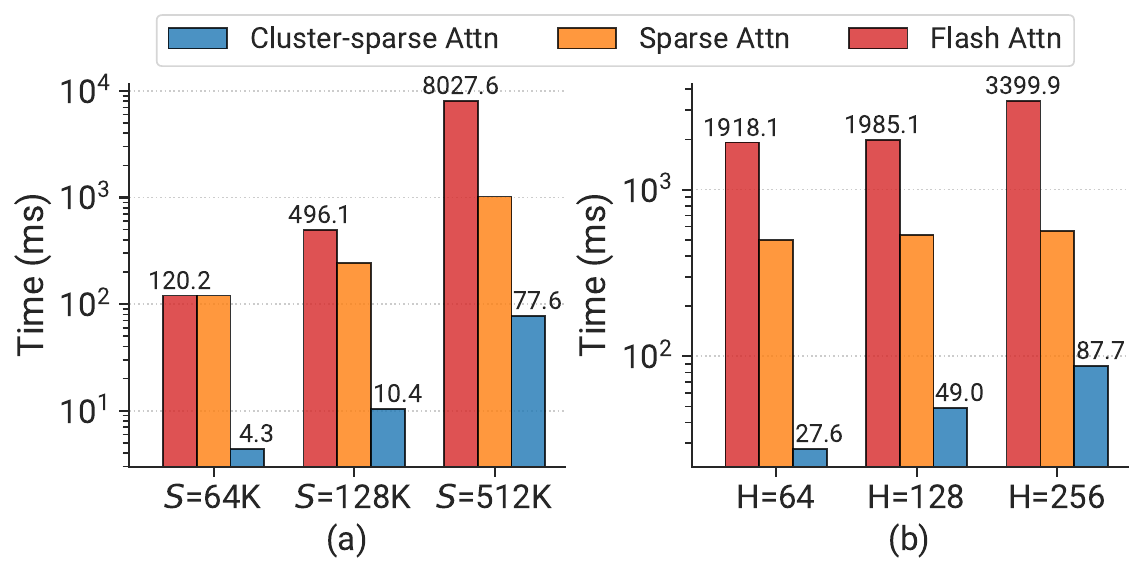}
    \caption{Computation time of attention modules when training Graphormer on ogbn-products with different (a) sequence lengths and (b) model hidden dimensions when $S$=256K.}
    \label{fig_attn_time}
\end{figure}

\textit{(1) Attention computation time w.r.t. sequence length.}
In Figure \ref{fig_attn_time}(a), we first evaluate the speed of the attention module when varying the sequence length from 64K to 512K. We use the model hyperparameter setting in GPH$_{Slim}$ and record the computation time of the attention module in each method. Clearly, as the sequence length increases, the computation time of FlashAttention grows quadratically, resulting in heavy training slowdown. The sparse attention improves the computation speed a bit, but shares a similar computation speed as FlashAttention when the sequence length is small. In contrast, \SysName essentially improves the computation efficiency by up to 103.4$\times$ compared to FlashAttention. It is even faster than the sparse attention largely, validating its effectiveness in reducing irregular memory access with our cluster-sparse pattern.

\textit{(2) Attention computation time w.r.t. hidden dimension.}
We fix the sequence length to 256K and change the hidden dimension from 64 to 256. The computation time of the attention module is recorded in Figure \ref{fig_attn_time}(b). When the model size increases, \SysName still largely outperforms FlashAttention and sparse attention in all cases, owing to the specialized cluster sparsity in \textit{Elastic Computation Reformation} module. We can conclude that FlashAttention shows poorer adaptation on long sequences, compared to its higher tolerance on larger model sizes. This indicates the better scalability of \SysName for long sequences and large model sizes.

\textit{(3) Sensitivity analysis of transfer threshold.} 
The introduced hyperparameter $\beta_{thre}$ determines the model performance and efficiency of \SysName. As shown in Table \ref{table_ablation_p}, we adopt different values of $\beta_{thre}$ and record the training time per epoch and test accuracy. Note that a larger $\beta_{thre}$ means more clusters are transferred to dense ones with sub-blocks. Higher accuracy can be obtained when smaller $\beta_{thre}$ is adopted, but coming with possibly lower training speed (e.g., for GraphSAGE, 0.368s with $\beta_{thre}=\beta_{G}$ vs. 0.077s with $\beta_{thre}=10\beta_{G}$). Seriously chasing the lowest error ($\beta_{thre}=0$) or just caring about the highest training throughput ($\beta_{thre}=1$) is not the best choice to fully utilize the benefits of the cluster-sparse pattern. Choosing different values always creates a trade-off between efficiency and accuracy and further analysis on the value choice can be done in the future. Currently, we suggest $\beta_{thre}=5\beta_{G}$ for better balance.

\begin{table}[t]
    \centering
    \caption{Training time per epoch and test accuracy on ogbn-arxiv dataset regarding different transfer threshold $\beta_{thre}$.}
    \resizebox{\linewidth}{!}{
      \begin{tabular}{@{}cccccccc@{}}
\toprule
\multicolumn{2}{c}{$\beta_{thre}$}                         & $\beta_{G}$   & 1.5$\beta_{G}$ & 5$\beta_{G}$  & 7$\beta_{G}$  & 10$\beta_{G}$ & \SysName \\ \midrule
\multirow{2}{*}{\textbf{GPH$_{Slim}$}} & \textbf{$t_{epoch}$(s)}  & 0.368 & 0.257  & 0.088 & 0.087 & 0.077 & 0.114    \\
                                       & \textbf{Test   Acc.(\%)} & 53.34 & 54.19  & 53.82 & 50.84 & 48.31 & 53.81    \\ \midrule
\multirow{2}{*}{\textbf{GT}}           & \textbf{$t_{epoch}$(s)}  & 0.098 & 0.090  & 0.089 & 0.084 & 0.071 & 0.093    \\
                                       & \textbf{Test   Acc.(\%)} & 56.70 & 56.84  & 56.51 & 53.65 & 45.95 & 56.72    \\ \bottomrule
\end{tabular}
            }
    \label{table_ablation_p}
\end{table}

\subsection{Pre-processing Cost} 
We record the pre-processing cost versus model convergence time on both tasks to understand how much extra time is brought by \SysName. The proportion is 5.2s (5.4\%) versus 91.2s (94.6\%) for ogbn-arxiv, and 239.7s (2.0\%) versus 11732.4s (98.0\%) for MalNet. The overhead only occupies less than 5.4\% of the total training time on all epochs, which is acceptable compared with the huge model convergence time.




%% file: 5_Related_Work.tex
\section{Related Work}
\label{sec_related}

As a new kind of graph learning algorithms outperforming traditional GNNs, many graph transformer architectures have arisen in recent years. Among them, models \cite{Graphormer,NAGphormer,Equiformer,smallgraph-case1,smallgraph-case2} utilize standard attention as foundation encoders to capture the all-pair interactions between nodes, leading to quadratic computation complexity. Some other works \cite{NAGphormer,Gophormer,Gapformer} adopt sampling or pooling methods which only select a subset of nodes to be trained at each iteration, without reducing the computation complexity. \cite{SGFormer,NodeFormer,DIFFormer} use self-defined adapted attention with poor generality and scalability.
As for system optimizations for transformers, sparse attention \cite{Performer,BIGBIRD,Linformer,Reformer} has been widely studied in NLP area for linear complexity. Borrowing this, \cite{Exphormer,GraphGPS,GT} directly apply graph topology in the attention computation. Besides, multiple works for LLM \cite{sp-colossalai,sp-ds,sp-megatron,BlockwiseRingAttn} split the input sentence sequences and train distributedly for larger scalability.

%% file: 6_Conclusion.tex
\section{Conclusion}
\label{sec_conclusion}
To conclude, \SysName reveals challenges and opportunities in training graph transformer on large graphs, with our efforts in designing a scalable and efficient training system.


%% file: sc24.bbl
\begin{thebibliography}{10}
\providecommand{\url}[1]{#1}
\csname url@samestyle\endcsname
\providecommand{\newblock}{\relax}
\providecommand{\bibinfo}[2]{#2}
\providecommand{\BIBentrySTDinterwordspacing}{\spaceskip=0pt\relax}
\providecommand{\BIBentryALTinterwordstretchfactor}{4}
\providecommand{\BIBentryALTinterwordspacing}{\spaceskip=\fontdimen2\font plus
\BIBentryALTinterwordstretchfactor\fontdimen3\font minus \fontdimen4\font\relax}
\providecommand{\BIBforeignlanguage}[2]{{%
\expandafter\ifx\csname l@#1\endcsname\relax
\typeout{** WARNING: IEEEtran.bst: No hyphenation pattern has been}%
\typeout{** loaded for the language `#1'. Using the pattern for}%
\typeout{** the default language instead.}%
\else
\language=\csname l@#1\endcsname
\fi
#2}}
\providecommand{\BIBdecl}{\relax}
\BIBdecl

\bibitem{GCN}
T.~N. Kipf and M.~Welling, ``Semi-supervised classification with graph convolutional networks,'' in \emph{International Conference on Learning Representations}, ser. ICLR '16, 2016.

\bibitem{GraphSAGE}
W.~Hamilton, Z.~Ying, and J.~Leskovec, ``Inductive representation learning on large graphs,'' in \emph{Advances in Neural Information Processing Systems}, ser. NeurIPS '17, 2017.

\bibitem{GAT}
P.~Veličković, G.~Cucurull, A.~Casanova, A.~Romero, P.~Liò, and Y.~Bengio, ``Graph attention networks,'' in \emph{International Conference on Learning Representations}, ser. ICLR '18, 2018.

\bibitem{PowerfulGNN}
K.~Xu, W.~Hu, J.~Leskovec, and S.~Jegelka, ``How powerful are graph neural networks?'' \emph{CoRR}, vol. abs/1810.00826, 2019.

\bibitem{PUFFIN}
P.~Li, Y.~Guo, Y.~Luo, X.~Wang, Z.~Wang, and X.~Liu, ``Graph neural networks based memory inefficiency detection using selective sampling,'' in \emph{Proceedings of the International Conference on High Performance Computing, Networking, Storage and Analysis}, ser. SC '22.\hskip 1em plus 0.5em minus 0.4em\relax IEEE Press, 2022.

\bibitem{CoGNN}
Q.~Sun, Y.~Liu, H.~Yang, R.~Zhang, M.~Dun, M.~Li, X.~Liu, W.~Xiao, Y.~Li, Z.~Luan, and D.~Qian, ``Cognn: efficient scheduling for concurrent gnn training on gpus,'' in \emph{Proceedings of the International Conference on High Performance Computing, Networking, Storage and Analysis}, ser. SC '22.\hskip 1em plus 0.5em minus 0.4em\relax IEEE Press, 2022.

\bibitem{Sylvie}
M.~Zhang, Q.~Hu, C.~Wan, H.~Wang, P.~Sun, Y.~Wen, and T.~Zhang, ``Sylvie: 3d-adaptive and universal system for large-scale graph neural network training,'' in \emph{2024 IEEE 38th International Conference on Data Engineering (ICDE)}, 2024.

\bibitem{GNN-survey}
S.~Abadal, A.~Jain, R.~Guirado, J.~López-Alonso, and E.~Alarcón, ``Computing graph neural networks: A survey from algorithms to accelerators,'' \emph{CoRR}, vol. abs/2010.00130, 2021.

\bibitem{GNNOverSmooth}
D.~Chen, Y.~Lin, W.~Li, P.~Li, J.~Zhou, and X.~Sun, ``Measuring and relieving the over-smoothing problem for graph neural networks from the topological view,'' in \emph{Proceedings of the AAAI Conference on Artificial Intelligence}, ser. AAAI '20, 2020.

\bibitem{GNNOverSquash}
U.~Alon and E.~Yahav, ``On the bottleneck of graph neural networks and its practical implications,'' in \emph{International Conference on Learning Representations}, ser. ICLR '21, 2021.

\bibitem{GNNOverSquash22}
J.~Topping, F.~D. Giovanni, B.~P. Chamberlain, X.~Dong, and M.~M. Bronstein, ``Understanding over-squashing and bottlenecks on graphs via curvature,'' in \emph{International Conference on Learning Representations}, ser. ICLR '22, 2022.

\bibitem{HighOrderGNN}
C.~Morris, M.~Ritzert, M.~Fey, W.~L. Hamilton, J.~E. Lenssen, G.~Rattan, and M.~Grohe, ``Weisfeiler and leman go neural: Higher-order graph neural networks,'' in \emph{Proceedings of the AAAI Conference on Artificial Intelligence}, ser. AAAI '19, 2019.

\bibitem{Attention}
A.~Vaswani, N.~Shazeer, N.~Parmar, J.~Uszkoreit, L.~Jones, A.~N. Gomez, Å.~Kaiser, and I.~Polosukhin, ``Attention is all you need,'' in \emph{Advances in Neural Information Processing Systems}, ser. NeurIPS '17, 2017.

\bibitem{Graphormer}
C.~Ying, T.~Cai, S.~Luo, S.~Zheng, G.~Ke, D.~He, Y.~Shen, and T.-Y. Liu, ``Do transformers really perform badly for graph representation?'' in \emph{Advances in Neural Information Processing Systems}, ser. NeurIPS '21, 2021-12-06.

\bibitem{GT}
V.~P. Dwivedi and X.~Bresson, ``A generalization of transformer networks to graphs,'' \emph{CoRR}, vol. abs/2012.09699, 2021.

\bibitem{NAGphormer}
J.~Chen, K.~Gao, G.~Li, and K.~He, ``Nagphormer: A tokenized graph transformer for node classification in large graphs,'' in \emph{International Conference on Learning Representations}, ser. ICLR '23, 2023.

\bibitem{NodeFormer}
Q.~Wu, W.~Zhao, Z.~Li, D.~Wipf, and J.~Yan, ``Nodeformer: A scalable graph structure learning transformer for node classification,'' in \emph{Advances in Neural Information Processing Systems}, ser. NeurIPS '22, 2022.

\bibitem{SAN}
D.~Kreuzer, D.~Beaini, W.~L. Hamilton, V.~Létourneau, and P.~Tossou, ``Rethinking graph transformers with spectral attention,'' in \emph{Advances in Neural Information Processing Systems}, ser. NeurIPS '21, 2021.

\bibitem{SAT}
D.~Chen, L.~O'Bray, and K.~Borgwardt, ``Structure-aware transformer for graph representation learning,'' in \emph{Proceedings of the 39th International Conference on Machine Learning}, ser. ICML '22, 2022, pp. 3469--3489.

\bibitem{DIFFormer}
Q.~Wu, C.~Yang, W.~Zhao, Y.~He, D.~Wipf, and J.~Yan, ``Difformer: Scalable (graph) transformers induced by energy constrained diffusion,'' in \emph{International Conference on Learning Representations}, ser. ICLR '23, 2023.

\bibitem{SGFormer}
Q.~Wu, W.~Zhao, C.~Yang, H.~Zhang, F.~Nie, H.~Jiang, Y.~Bian, and J.~Yan, ``Simplifying and empowering transformers for large-graph representations,'' in \emph{Advances in Neural Information Processing Systems}, ser. NeurIPS '23, 2023.

\bibitem{GraphTrans}
Z.~Wu, P.~Jain, M.~Wright, A.~Mirhoseini, J.~E. Gonzalez, and I.~Stoica, ``Representing long-range context for graph neural networks with global attention,'' in \emph{Advances in Neural Information Processing Systems}, ser. NeurIPS '21, 2021-12-06.

\bibitem{OGB}
W.~Hu, M.~Fey, M.~Zitnik, Y.~Dong, H.~Ren, B.~Liu, M.~Catasta, and J.~Leskovec, ``Open graph benchmark: Datasets for machine learning on graphs,'' in \emph{Advances in Neural Information Processing Systems}, ser. NeurIPS '20, 2020.

\bibitem{amazon}
R.~He and J.~McAuley, ``Ups and downs: Modeling the visual evolution of fashion trends with one-class collaborative filtering,'' in \emph{Proceedings of the 25th International Conference on World Wide Web}, ser. WWW '16, 2016.

\bibitem{GraphGPS}
L.~Rampášek, M.~Galkin, V.~P. Dwivedi, A.~T. Luu, G.~Wolf, and D.~Beaini, ``Recipe for a general, powerful, scalable graph transformer,'' in \emph{Advances in Neural Information Processing Systems}, ser. NeurIPS '22, 2022-12-06.

\bibitem{Exphormer}
H.~Shirzad, A.~Velingker, B.~Venkatachalam, D.~J. Sutherland, and A.~K. Sinop, ``Exphormer: Sparse transformers for graphs,'' in \emph{Proceedings of the 40th International Conference on Machine Learning}, ser. ICML '23, 2023, pp. 31\,613--31\,632.

\bibitem{smallgraph-case1}
L.~Chanussot, A.~Das, S.~Goyal, T.~Lavril, M.~Shuaibi, M.~Riviere, K.~Tran, J.~Heras-Domingo, C.~Ho, W.~Hu, A.~Palizhati, A.~Sriram, B.~Wood, J.~Yoon, D.~Parikh, C.~L. Zitnick, and Z.~Ulissi, ``The open catalyst 2020 (oc20) dataset and community challenges,'' \emph{CoRR}, vol. abs/2010.09990, 2021.

\bibitem{smallgraph-case2}
Z.~Fan, T.~Chen, P.~Wang, and Z.~Wang, ``Cadtransformer: Panoptic symbol spotting transformer for cad drawings,'' in \emph{Proceedings of the IEEE/CVF Conference on Computer Vision and Pattern Recognition}, 2022, pp. 10\,986--10\,996.

\bibitem{NI-CTR}
E.~Min, Y.~Rong, T.~Xu, Y.~Bian, P.~Zhao, J.~Huang, D.~Luo, K.~Lin, and S.~Ananiadou, ``Neighbour interaction based click-through rate prediction via graph-masked transformer,'' \emph{CoRR}, vol. abs/2201.13311, 2022.

\bibitem{Equiformer}
Y.-L. Liao and T.~Smidt, ``Equiformer: Equivariant graph attention transformer for 3d atomistic graphs,'' in \emph{International Conference on Learning Representations}, ser. ICLR '23, 2023.

\bibitem{FlashAttention}
T.~Dao, D.~Fu, S.~Ermon, A.~Rudra, and C.~Ré, ``Flashattention: Fast and memory-efficient exact attention with io-awareness,'' in \emph{Advances in Neural Information Processing Systems}, ser. NeurIPS '22, 2022-12-06.

\bibitem{Gophormer}
J.~Zhao, C.~Li, Q.~Wen, Y.~Wang, Y.~Liu, H.~Sun, X.~Xie, and Y.~Ye, ``Gophormer: Ego-graph transformer for node classification,'' \emph{CoRR}, vol. abs/2110.13094, 2021.

\bibitem{FastGCN}
J.~Chen, T.~Ma, and C.~Xiao, ``Fastgcn: Fast learning with graph convolutional networks via importance sampling,'' in \emph{International Conference on Learning Representations}, ser. ICLR '18, 2018.

\bibitem{GraphSAINT}
H.~Zeng, H.~Zhou, A.~Srivastava, R.~Kannan, and V.~Prasanna, ``Graphsaint: Graph sampling based inductive learning method,'' in \emph{International Conference on Learning Representations}, ser. ICLR '20, 2020.

\bibitem{Performer}
K.~M. Choromanski, V.~Likhosherstov, D.~Dohan, X.~Song, A.~Gane, T.~Sarlos, P.~Hawkins, J.~Q. Davis, A.~Mohiuddin, L.~Kaiser, D.~B. Belanger, L.~J. Colwell, and A.~Weller, ``Rethinking attention with performers,'' in \emph{International Conference on Learning Representations}, ser. ICLR '23, 2023.

\bibitem{BIGBIRD}
M.~Zaheer, G.~Guruganesh, K.~A. Dubey, J.~Ainslie, C.~Alberti, S.~Ontanon, P.~Pham, A.~Ravula, Q.~Wang, L.~Yang \emph{et~al.}, ``Big bird: Transformers for longer sequences,'' \emph{Advances in neural information processing systems}, vol.~33, pp. 17\,283--17\,297, 2020.

\bibitem{sp-colossalai}
\BIBentryALTinterwordspacing
S.~Li, F.~Xue, C.~Baranwal, Y.~Li, and Y.~You, ``Sequence parallelism: Long sequence training from system perspective,'' in \emph{Proceedings of the 61st Annual Meeting of the Association for Computational Linguistics (Volume 1: Long Papers)}, A.~Rogers, J.~Boyd-Graber, and N.~Okazaki, Eds.\hskip 1em plus 0.5em minus 0.4em\relax Toronto, Canada: Association for Computational Linguistics, Jul. 2023, pp. 2391--2404. [Online]. Available: \url{https://aclanthology.org/2023.acl-long.134}
\BIBentrySTDinterwordspacing

\bibitem{sp-ds}
S.~A. Jacobs, M.~Tanaka, C.~Zhang, M.~Zhang, S.~L. Song, S.~Rajbhandari, and Y.~He, ``Deepspeed ulysses: System optimizations for enabling training of extreme long sequence transformer models,'' \emph{CoRR}, vol. abs/2309.14509, 2023.

\bibitem{sp-megatron}
V.~A. Korthikanti, J.~Casper, S.~Lym, L.~McAfee, M.~Andersch, M.~Shoeybi, and B.~Catanzaro, ``Reducing activation recomputation in large transformer models,'' in \emph{Proceedings of Machine Learning and Systems}, ser. MLSys '23, 2023.

\bibitem{BlockwiseRingAttn}
H.~Liu, M.~Zaharia, and P.~Abbeel, ``Ring attention with blockwise transformers for near-infinite context,'' \emph{CoRR}, vol. abs/2310.01889, 2023.

\bibitem{Mizan}
Z.~Khayyat, K.~Awara, A.~Alonazi, H.~Jamjoom, D.~Williams, and P.~Kalnis, ``Mizan: a system for dynamic load balancing in large-scale graph processing,'' in \emph{Proceedings of the 8th ACM European Conference on Computer Systems}, ser. EuroSys '13.\hskip 1em plus 0.5em minus 0.4em\relax Association for Computing Machinery, 2013.

\bibitem{Enterprise}
H.~Liu and H.~H. Huang, ``Enterprise: breadth-first graph traversal on gpus,'' in \emph{Proceedings of the International Conference for High Performance Computing, Networking, Storage and Analysis}, ser. SC '15.\hskip 1em plus 0.5em minus 0.4em\relax Association for Computing Machinery, 2015.

\bibitem{Powerlaw}
A.~Sala, H.~Zheng, B.~Y. Zhao, S.~Gaito, and G.~P. Rossi, ``Brief announcement: revisiting the power-law degree distribution for social graph analysis,'' in \emph{Proceedings of the 29th ACM SIGACT SIGOPS symposium on Principles of distributed computing}, ser. PODC '10, 2010.

\bibitem{GPT3}
T.~Brown, B.~Mann, N.~Ryder, M.~Subbiah, J.~D. Kaplan, P.~Dhariwal, A.~Neelakantan, P.~Shyam, G.~Sastry, A.~Askell, S.~Agarwal, A.~Herbert-Voss, G.~Krueger, T.~Henighan, R.~Child, A.~Ramesh, D.~Ziegler, J.~Wu, C.~Winter, C.~Hesse, M.~Chen, E.~Sigler, M.~Litwin, S.~Gray, B.~Chess, J.~Clark, C.~Berner, S.~McCandlish, A.~Radford, I.~Sutskever, and D.~Amodei, ``Language models are few-shot learners,'' in \emph{Advances in Neural Information Processing Systems}, ser. NeurIPS '20, 2020.

\bibitem{LLaMA}
H.~Touvron, T.~Lavril, G.~Izacard, X.~Martinet, M.-A. Lachaux, T.~Lacroix, B.~Rozière, N.~Goyal, E.~Hambro, F.~Azhar, A.~Rodriguez, A.~Joulin, E.~Grave, and G.~Lample, ``Llama: Open and efficient foundation language models,'' \emph{CoRR}, vol. abs/2302.13971, 2023.

\bibitem{BlocksparseRNN}
S.~Narang, E.~Undersander, and G.~Diamos, ``Block-sparse recurrent neural networks,'' \emph{CoRR}, vol. abs/1711.02782, 2017.

\bibitem{BlockBert}
J.~Qiu, H.~Ma, O.~Levy, W.-t. Yih, S.~Wang, and J.~Tang, ``Blockwise self-attention for long document understanding,'' in \emph{Findings of the Association for Computational Linguistics: EMNLP 2020}, 2020.

\bibitem{AutoGT}
Z.~Zhang, X.~Wang, C.~Guan, Z.~Zhang, H.~Li, and W.~Zhu, ``Autogt: Automated graph transformer architecture search,'' in \emph{International Conference on Learning Representations}, ser. ICLR '23, 2023.

\bibitem{EGT}
M.~S. Hussain, M.~J. Zaki, and D.~Subramanian, ``Global self-attention as a replacement for graph convolution,'' \emph{CoRR}, vol. abs/2108.03348, 2022.

\bibitem{GraphBert}
J.~Zhang, H.~Zhang, C.~Xia, and L.~Sun, ``Graph-bert: Only attention is needed for learning graph representations,'' \emph{CoRR}, vol. abs/2001.05140, 2020.

\bibitem{UniMP}
\BIBentryALTinterwordspacing
Y.~Shi, Z.~Huang, S.~Feng, H.~Zhong, W.~Wang, and Y.~Sun, ``Masked label prediction: Unified message passing model for semi-supervised classification,'' in \emph{Proceedings of the Thirtieth International Joint Conference on Artificial Intelligence, {IJCAI-21}}, Z.-H. Zhou, Ed.\hskip 1em plus 0.5em minus 0.4em\relax International Joint Conferences on Artificial Intelligence Organization, 8 2021, pp. 1548--1554, main Track. [Online]. Available: \url{https://doi.org/10.24963/ijcai.2021/214}
\BIBentrySTDinterwordspacing

\bibitem{PLAN}
L.~M.~S. Khoo, H.~L. Chieu, Z.~Qian, and J.~Jiang, ``Interpretable rumor detection in microblogs by attending to user interactions,'' in \emph{Proceedings of the AAAI Conference on Artificial Intelligence}, ser. AAAI '20, 2020.

\bibitem{Malnet}
S.~Freitas, Y.~Dong, J.~Neil, and D.~H. Chau, ``A large-scale database for graph representation learning,'' \emph{arXiv preprint arXiv:2011.07682}, 2020.

\bibitem{SpAtten}
\BIBentryALTinterwordspacing
H.~Wang, Z.~Zhang, and S.~Han, ``Spatten: Efficient sparse attention architecture with cascade token and head pruning,'' in \emph{2021 IEEE International Symposium on High-Performance Computer Architecture (HPCA)}.\hskip 1em plus 0.5em minus 0.4em\relax Los Alamitos, CA, USA: IEEE Computer Society, mar 2021, pp. 97--110. [Online]. Available: \url{https://doi.ieeecomputersociety.org/10.1109/HPCA51647.2021.00018}
\BIBentrySTDinterwordspacing

\bibitem{Linformer}
S.~Wang, B.~Z. Li, M.~Khabsa, H.~Fang, and H.~Ma, ``Linformer: Self-attention with linear complexity,'' \emph{CoRR}, vol. abs/2006.04768, 2020.

\bibitem{Reformer}
N.~Kitaev, L.~Kaiser, and A.~Levskaya, ``Reformer: The efficient transformer,'' in \emph{International Conference on Learning Representations}, ser. ICLR '20, 2020.

\bibitem{Gapformer}
\BIBentryALTinterwordspacing
C.~Liu, Y.~Zhan, X.~Ma, L.~Ding, D.~Tao, J.~Wu, and W.~Hu, ``Gapformer: Graph transformer with graph pooling for node classification,'' in \emph{Proceedings of the Thirty-Second International Joint Conference on Artificial Intelligence, {IJCAI-23}}, E.~Elkind, Ed.\hskip 1em plus 0.5em minus 0.4em\relax International Joint Conferences on Artificial Intelligence Organization, 8 2023, pp. 2196--2205, main Track. [Online]. Available: \url{https://doi.org/10.24963/ijcai.2023/244}
\BIBentrySTDinterwordspacing

\bibitem{TransformerApproximability}
C.~Yun, Y.-W. Chang, S.~Bhojanapalli, A.~S. Rawat, S.~Reddi, and S.~Kumar, ``O(n) connections are expressive enough: Universal approximability of sparse transformers,'' \emph{Advances in Neural Information Processing Systems}, vol.~33, pp. 13\,783--13\,794, 2020.

\bibitem{Hamiltonian}
``Hamiltonian path,'' \url{https://mathworld.wolfram.com/HamiltonianPath.html}, 2020.

\bibitem{Dirac}
\BIBentryALTinterwordspacing
C.~Lee and B.~Sudakov, ``Dirac's theorem for random graphs,'' \emph{CoRR}, vol. abs/1108.2502v3, 2012. [Online]. Available: \url{https://arxiv.org/abs/1108.2502}
\BIBentrySTDinterwordspacing

\bibitem{Community1}
S.~Fortunato, ``Community detection in graphs,'' \emph{Physics Reports}, vol. 486, no.~3, pp. 75--174, 2010.

\bibitem{SpectralCommunityDetect}
\BIBentryALTinterwordspacing
M.~E.~J. Newman, ``Spectral methods for community detection and graph partitioning,'' \emph{Phys. Rev. E}, vol.~88, p. 042822, Oct 2013. [Online]. Available: \url{https://link.aps.org/doi/10.1103/PhysRevE.88.042822}
\BIBentrySTDinterwordspacing

\bibitem{GraphPartition}
``Graph partitioning models for parallel computing,'' \emph{Parallel Computing}, vol.~26, no.~12, pp. 1519--1534, 2000, graph Partitioning and Parallel Computing.

\bibitem{GNNAdvisor}
Y.~Wang, B.~Feng, G.~Li, S.~Li, L.~Deng, Y.~Xie, and Y.~Ding, ``{GNNAdvisor}: An adaptive and efficient runtime system for {GNN} acceleration on {GPUs},'' in \emph{15th USENIX Symposium on Operating Systems Design and Implementation}, ser. OSDI '21.\hskip 1em plus 0.5em minus 0.4em\relax USENIX Association, 2021, pp. 515--531.

\bibitem{METIS}
K.~George and K.~Vipin, ``Metis: A software package for partitioning unstructured graphs, partitioning meshes, and computing fill-reducing orderings of sparse matrices,'' 1997.

\bibitem{zinc}
V.~P. Dwivedi, C.~K. Joshi, A.~T. Luu, T.~Laurent, Y.~Bengio, and X.~Bresson, ``Benchmarking graph neural networks,'' \emph{Journal of Machine Learning Research}, vol.~24, no.~43, pp. 1--48, 2023.

\bibitem{PyTorch}
A.~Paszke, S.~Gross, F.~Massa, A.~Lerer, J.~Bradbury, G.~Chanan, T.~Killeen, Z.~Lin, N.~Gimelshein, L.~Antiga, A.~Desmaison, A.~Kopf, E.~Yang, Z.~DeVito, M.~Raison, A.~Tejani, S.~Chilamkurthy, B.~Steiner, L.~Fang, J.~Bai, and S.~Chintala, ``Pytorch: An imperative style, high-performance deep learning library,'' in \emph{Advances in Neural Information Processing Systems}, ser. NeurIPS '19, 2019.

\end{thebibliography}
